\documentclass[fleqn,usenatbib]{mnras}

\usepackage[T1]{fontenc}
\usepackage{ae,aecompl}

\usepackage{graphicx}	% Including figure files
\usepackage{amsmath}	% Advanced maths commands
\usepackage{amssymb}	% Extra maths symbols
\usepackage{float}
\title[IR excesses in stars with and without planets]{Infrared excesses in stars with and without planets 
using revised {\it WISE} photometry}

\author[R. F. Maldonado et al.]{
Raul F. Maldonado,\thanks{E-mail: raulfms@inaoep.mx}
Miguel Chavez, Emanuele Bertone, and \newauthor
Fernando Cruz-Saenz de Miera\\
Instituto Nacional de Astrofisica, Optica y Electronica, Luis Enrique Erro 1, Tonantzintla, Puebla, Mexico\\
}
\date{Accepted XXX. Received YYY; in original form ZZZ}

\pubyear{2017}

\begin{document}
\label{firstpage}
\pagerange{\pageref{firstpage}--\pageref{lastpage}}
\maketitle

% Abstract of the paper
\begin{abstract}
We present an analysis on the potential prevalence of mid infrared excesses in stars with and without planetary companions. 
 Based on an extended database of stars detected with the {\it WISE} satellite, we studied two stellar samples: one with 236 planet hosts and another with 986 objects for which planets have been searched but not found. We determined the presence of an excess over the photosphere by comparing the observed flux ratio at 22~$\mu$m and 12~$\mu$m ($f_{22}/f_{12}$) with the corresponding synthetic value, derived from results of classical model photospheres.
We found a detection rate of 0.85$\%$ at 22~$\mu$m (2 excesses) in the sample of stars with planets and 0.1$\%$ (1 detection) for the stars without planets. The difference of the detection rate between the two samples is not statistically significant, a result that is independent of the different approaches found in the literature to define an excess in the wavelength range covered by {\it WISE} observations. As an additional result, we found that the {\it WISE} fluxes required a 
normalisation procedure  to make  them compatible with synthetic data, probably pointing out a revision of the {\it WISE} data calibration.

\end{abstract}

% Select between one and six entries from the list of approved keywords.
% Don't make up new ones.
\begin{keywords}
Stars: circumstellar matter -- planetary systems -- infrared: stars
\end{keywords}

%%%%%%%%%%%%%%%%%%%%%%%%%%%%%%%%%%%%%%%%%%%%%%%%%%

%%%%%%%%%%%%%%%%% BODY OF PAPER %%%%%%%%%%%%%%%%%%

\section{Introduction}

Discs and rings are common products of the evolution of different astronomical systems at different size scales, from galaxies 
to stars and planets. Circumstellar discs are found around stars at different evolutionary stages and they are commonly detected 
as an infrared (IR)  excess over the photospheric value, produced by their thermal emission. Debris discs represent the final 
stage of disc evolution, they may orbit mature systems and can coexist with a fully-formed planetary system, as in our solar case. 
They are formed by second-generation solid material, ranging from sub-mm dust particles to sub-planetary bodies, and, in a 
few cases, a small amount of gas has also been observed \citep[e.g.,][]{kospal2013}.
The grain size distribution is determined by the simultaneous processes that form (e.g., collisional grinding) or remove 
(e.g. Poynting-Robertson and stellar wind drag) solid particles \citep{wyatt2011}. Thus, this kind of disc continues to 
evolve collisionally and dynamically.

In the past decade, several surveys have explored debris discs around stars of spectral types from A to M in the mid- and far-IR, 
 probing dust at different locations, due to the positive correlation between the wavelength of the peak of the dust emission and the distance from the star\footnote{For a solar luminosity star, observations at 10~$\mu$m detect dust at a distance of $\sim$1~au, at 22~$\mu$m the distance is about 4.5~au, at 60~$\mu$m the distance increases to $\sim$33~au, while at 100~$\mu$m observations probe dust at about 90~au. These distances scale as the square root of the star luminosity.} \citep[e.g.,][]{wyatt2008}.
The most relevant surveys are FEPS\footnote{Formation and Evolution of Planetary Systems.} \citep{rieke2005}, using the {\em Spitzer} 
observatory,  DEBRIS \citep{sibthorpe2013} and DUNES\footnote{DUst around NEarby Stars.} \citep{eiroa2013}, carried out with the 
{\em Herschel} satellite.  \citet{matthews2014} made a comparison of detection rates of debris discs among the different 
surveys and found that, for A stars, the detection rates are 33$\%$ and 25$\%$ at 70~$\mu$m and 100~$\mu$m, respectively, 
while for solar type stars (FGK) it ranges from 10$\%$, in the case of FEPS survey, to 17$\%$, reported in the DEBRIS survey. 
DUNES increased the detection rate up to 20$\%$ \citep{eiroa2013}, while the recent work by \citet{montesinos2016} gives a value 
of 22\%. \citet{trilling2008} interpreted the apparent decrease of excess rates from A to K spectral types as due to an age effect, 
as expected  by circumstellar disc evolution. 

 A new kind of discs  was detected thanks to extensive surveys in the Mid-IR regime. These objects are called warm debris discs 
and the emission in excess found at wavelengths between 5~$\mu$m and 35~$\mu$m suggests the presence of material close to the star.  \citet{chen2005}, \citet{uzpen2007}, and \citet{hales2009} used {\it Spitzer} data, from 8 to 70~$\mu$m, to identify a significant number of flux excesses attributable to the presence of circumstellar warm dust.
More recent studies, that used  Wide Infrared Survey Explorer  ({\it WISE}) data, have found stars with IR excesses in 
the Mid-IR regime \citep{kennedy2012,cruz2014,patel2014,kuchner2016,costa2016}. We might expect that the presence of 
planets at few au from the stars (in the ``terrestrial" zone) and warm debris discs  are 
correlated, since planets are the final stage of agglomeration of planetesimals and the dust in warm debris discs is also 
formed from planetesimals. In this work we want to assess if the correlation between the Mid-IR excesses and planets does 
exist or not.  In this context, the work presented here complements previous analyses, in particular those of \citet{krivov2011} and 
\citet{morales2012}, who looked for IR excesses using, respectively, a preliminary {\it WISE} release and  the All-Sky {\it WISE} catalogue. 
While \citet{krivov2011} found no evidence for IR excesses in 52 transiting exoplanetary systems, \citet{morales2012}  found a 
prevalence of 2.6$\%$ in a large sample (591) of planetary systems around both main sequence (MS) and 
evolved stars; the incidence decreases to 1$\%$ if only MS stars are considered. At longer wavelengths (100~$\mu$m), \citet{moromartin2015} did not find significant difference in the occurrence of dust around stars hosting planets in selected samples of DUNES and DEBRIS objects. 

In this paper we analyze a sample of stars with planets and another comparison sample of stars where planets have been 
searched for but not found. Infrared data of the four {\it WISE} bands is used  to search for  IR excesses at 22~$\mu$m by comparing the observed and expected photospheric flux ratio ratio $f_{22}/f_{12}$. Furthermore, we also explore the 
issue by analysing different procedures used by other authors 
to obtain more results on the existence (or not) of a correlation 
between planets and warm debris discs \citep{morales2012,kennedy2012,patel2014,cruz2012}.

\section{{\it WISE}}

{\it WISE} was a NASA mission (2009-2011) whose main objective was to map the entire sky in four Mid-IR passbands at  
3.4~$\mu$m (W1), 4.6~$\mu$m (W2), 12~$\mu$m (W3) and 22~$\mu$m (W4). It consisted of a cryogenically cooled 
40~cm aperture telescope. It used four focal  planes that simultaneously took images of a 47$\times$47 arcmin field of view.  
The full-width-at-half- maximum (FWHM) of the point spread function (PSF) were 6.1'' for W1, 6.4\arcsec for 
W2, 6.5\arcsec for W3 and 12.0\arcsec for W4.

{\it WISE} photometry of point sources was conducted using PSF profile fitting and the background level was estimated in a 
ring at a fixed distance (form 50 to 70\arcsec) from the source. Two comprehensive photometric catalogues have been delivered. 
The first one, called All-Sky \citep{cutri2012},  covered more than 90$\%$ of the sky. The second data release, named 
All{\it WISE} \citep{cutri2013}, was delivered in November 2013 and included combined data of previous epochs of {\it WISE}, such as 
NEO{\it WISE}, which also provided photometry measurements of asteroids and comets, and Cryogenic {\it WISE},  which  obtained data  
before their cooling systems ended operations. Taking into account these combinations, the sensitivity and the 
photometric precision was improved as compared to the previous All-Sky catalogue.

{\it WISE} achieved sensitivities of $5\sigma$ for point sources with flux of 0.08, 0.11, 1 and 6~mJy (16.5, 15.5, 11.2 
and 7.9 Vega magnitudes, respectively) in the four bands, from W1 to W4. This  allowed to detect, in W1, faint 
objects that do not appear in the 2MASS $K_s$ survey, while at 12~$\mu$m and 25~$\mu$m the {\it WISE} survey is $\sim$100 
times more sensitive than IRAS \citep{neugebauer1984}.

\section{Homogeneising {\it WISE} and synthetic photometry.}
\label{sec:fluxcorr}

Prior to the search of IR excesses, we tested the agreement between the observed {\it WISE} 
fluxes, provided by the All{\it WISE} catalogue, and the ATLAS9 theoretical photospheric flux \citep{castelli2003} that 
we used as reference, using a very large sample of stars.

\subsection{The sample selection.}
\label{selectioncriteria}

We selected stars with spectral types from A0 to K9 and luminosity classes V, IV, or IV/V from the SIMBAD 
Astronomical Database\footnote{http://simbad.u-strasbg.fr/simbad/}. The Johnson V magnitude of these stars ranges 
from 5 to 12, which make them prone to have a good signal-to-noise ratio (S/N $\geq$ 5) in the {\it WISE} band W4. We 
then discarded all objects in multiple systems and pre-main sequence stars (T-Tauri, Ae/Be Herbig). 
We assembled an initial sample of 53\,349 objects.

We collected their {\it WISE} magnitudes from the All{\it WISE} catalogue, considering only objects that comply the following 
constraints: a) they must have a S/N$\geq$5 in W4 band; b) their W1, W2, W3 and W4 photometry must include an 
uncertainty measurement; c) they must be free from any artifact or contamination by nearby objects  (i.e. we excluded those sources considered as spurious detections or contaminated by diffraction spikes, scattered light halos or optical ghosts produced by a bright star on the same image, according to the criteria of allWISE catalogue); 
d) the saturation in W3 and W4 must be zero.

We also extracted the J, H, and $K_S$ magnitudes from the 2MASS All-Sky catalogue. The restrictions applied on this 
catalogue are as follows: a) J, H and $K_S$ photometry must be accompanied by the corresponding uncertainties; 
b) the quality flag of the photometry must be A or B, with a S/N$\geq$10 and S/N$\geq$7, respectively. 
Finally, we obtained the B and V photometry from SIMBAD. After applying the previous criteria, we obtained a 
sample of 24,117 stars.

\subsection{Correction for extinction and magnitude to flux conversion}

Due to their brightness range, we expect most of the stars in our sample to be nearby objects; however,  
we considered appropriate to apply a flux correction for those objects that show a considerable colour excess E(B-V). 
Correction for interstellar  extinction will reduce systematic errors in the determination of stellar parameters for 
which we use optical and near-IR photometry.

\begin{table}
\begin{center}
\caption{Zero magnitude flux density for B, V \citep{bessel1979}, J, H, $K_S$ (from http://ipac.caltech.edu/2mass), 
W1, W2, W3, W4 (from http://wise2.ipac.caltech.edu) bands.}
\label{zeropoint}
\begin{tabular}{l r@{}l}
\hline
Band & \multicolumn{2}{c}{\bf $F_0$} \\
   &  \multicolumn{2}{c}{(Jy)} \\
\hline
B (Johnson)	& 4266.&7 \\
V (Johnson)	& 3836.&3 \\
J (2MASS)	& 1594.&0 \\
H (2MASS)	& 1024.&0 \\
$K_S$ (2MASS)	&  666.&7 \\
W1 ({it WISE}g)	&   306.& 7 \\
W2 ({it WISE}g)	&  170.& 7 \\
W3 ({it WISE}g)	&    29.& 0 \\
W4 ({it WISE}g)	&   8.& 3\\
\hline
\end{tabular}
\end{center}
\end{table}

We first obtained the intrinsic colour $(B-V)_0$ for each object, based on its spectral type, from the colour calibration of 
 \citet{peacut2012}, and derived the colour excess $E(B-V)$. We computed the mean $\mu_{\rm E(B-V)}$ and the stardard deviation 
$\sigma_{\rm E(B-V)}$ of the $E(B-V)$ distribution, after applying a $3\sigma$-clipping to the data. We only applied a 
reddening correction, to all wavelength bands considered in this work, for the stars with 
$E(B-V)>\mu_{\rm E(B-V)} + 1 \sigma_{\rm E(B-V)}$= 0.078~mag. The mean value of the colour excess distribution is 0.026~mag.  
A total of 9230 stars were corrected for extinction through the equation:
\begin{equation}
f_\lambda = f_0 10^{ 0.4 (-m_\lambda+A_\lambda) } \, ,
\end{equation}
that also allows to convert apparent magnitudes $m_\lambda$ to fluxes $f_\lambda$. We used the 
zero magnitude flux $f_0$ in the Vega scale reported in Table~\ref{zeropoint} and the $A_\lambda$ values provided by 
the \citet{rieke1985} extinction curve. No colour correction was applied to {\it WISE} data, since the reddening at micron 
wavelengths is negligible.

\subsection{Photospheric fitting and synthetic photometry.}
\label{photfitting}

Since the search of IR excesses requires a good estimation of the photospheric contribution of each star, synthetic spectra are 
needed.  In the near and mid IR, the choice of a specific model atmosphere library is not critical:  \citet{sinclair2010} found 
that ATLAS9 \citep{castelli2003} and MARCS \citep{gustafson2008} present discrepancies smaller than $2\%$ at 24~$\mu$m for FGK stars, 
and \citet{cruz2012} obtain similar figures comparing ATLAS9 and NEXTGEN SEDs. The stellar spectra in the selected parameter space 
of the sample are well reproduced by the LTE plane-parallel, low resolution ATLAS9  model atmospheres 
\citep[see, e.g.,][]{bertone2004,martins2007}. In order to compute the synthetic photometry that we needed to determine the 
photospheric fluxes, we first estimated the atmospheric parameters of each star through  a match with a grid of ATLAS9 library.

\begin{figure*}
\begin{center}
\begin{tabular}{cc}
\includegraphics[width=9cm]{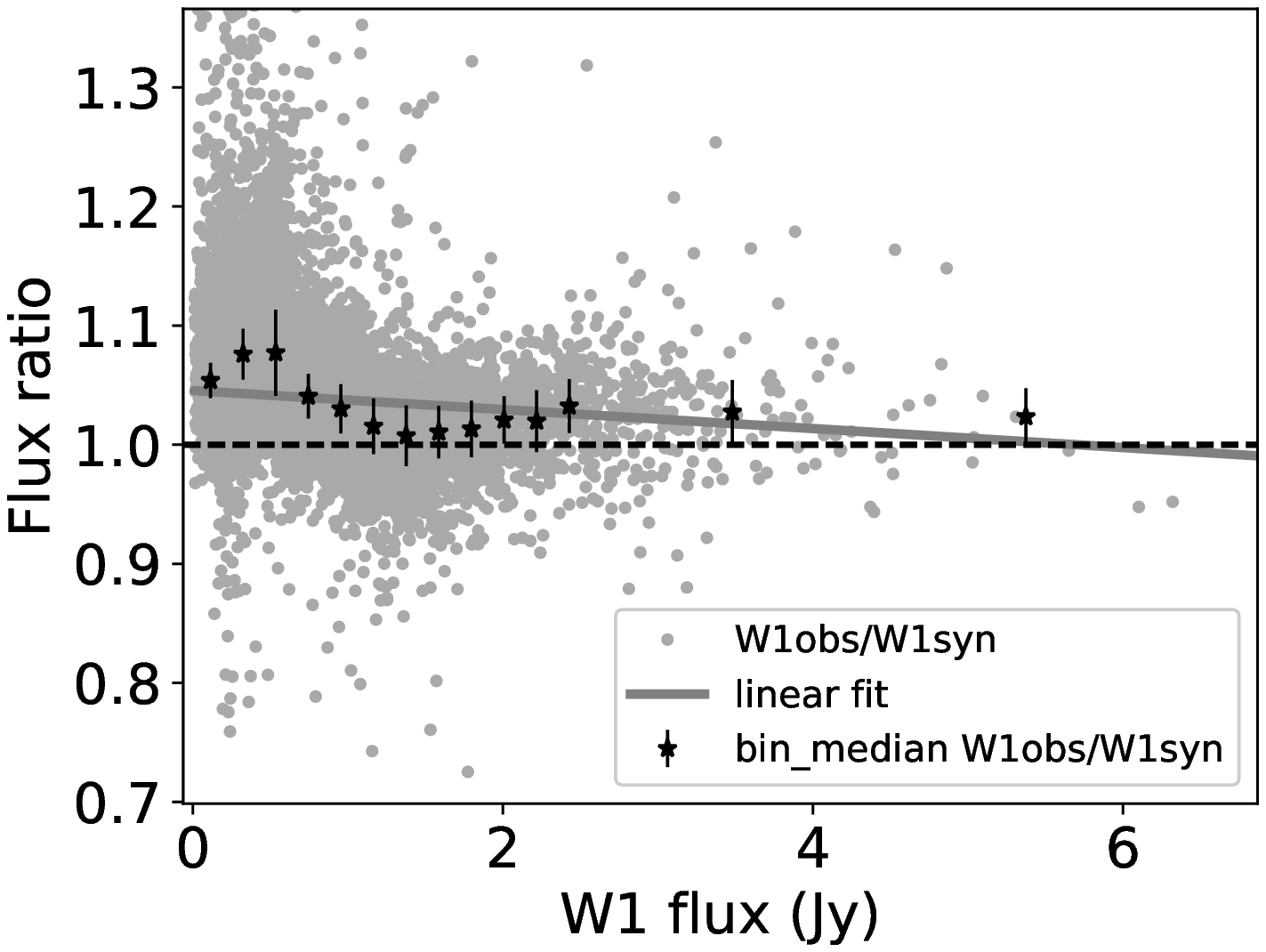} &
\includegraphics[width=9cm]{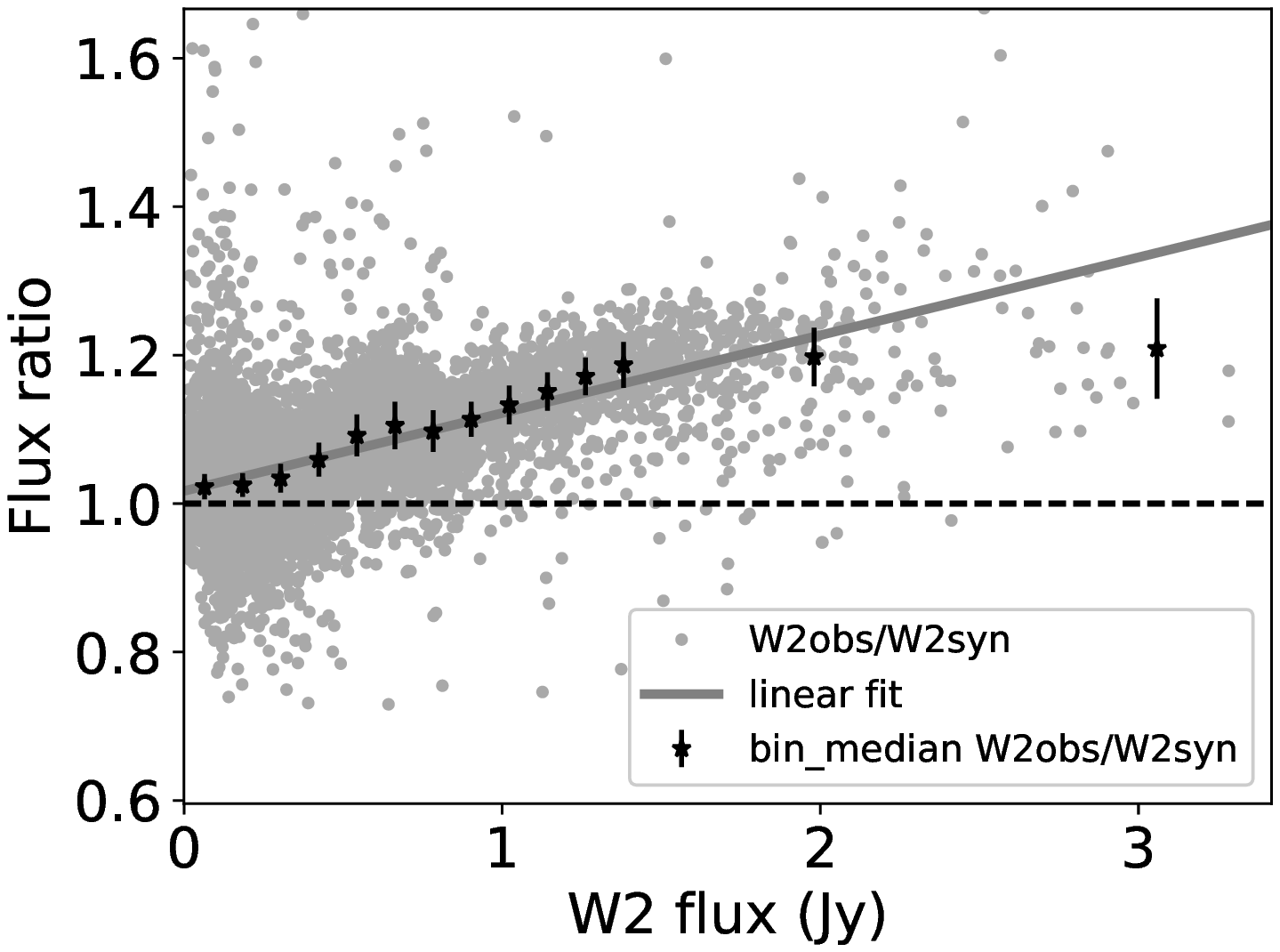} \\
\includegraphics[width=9cm]{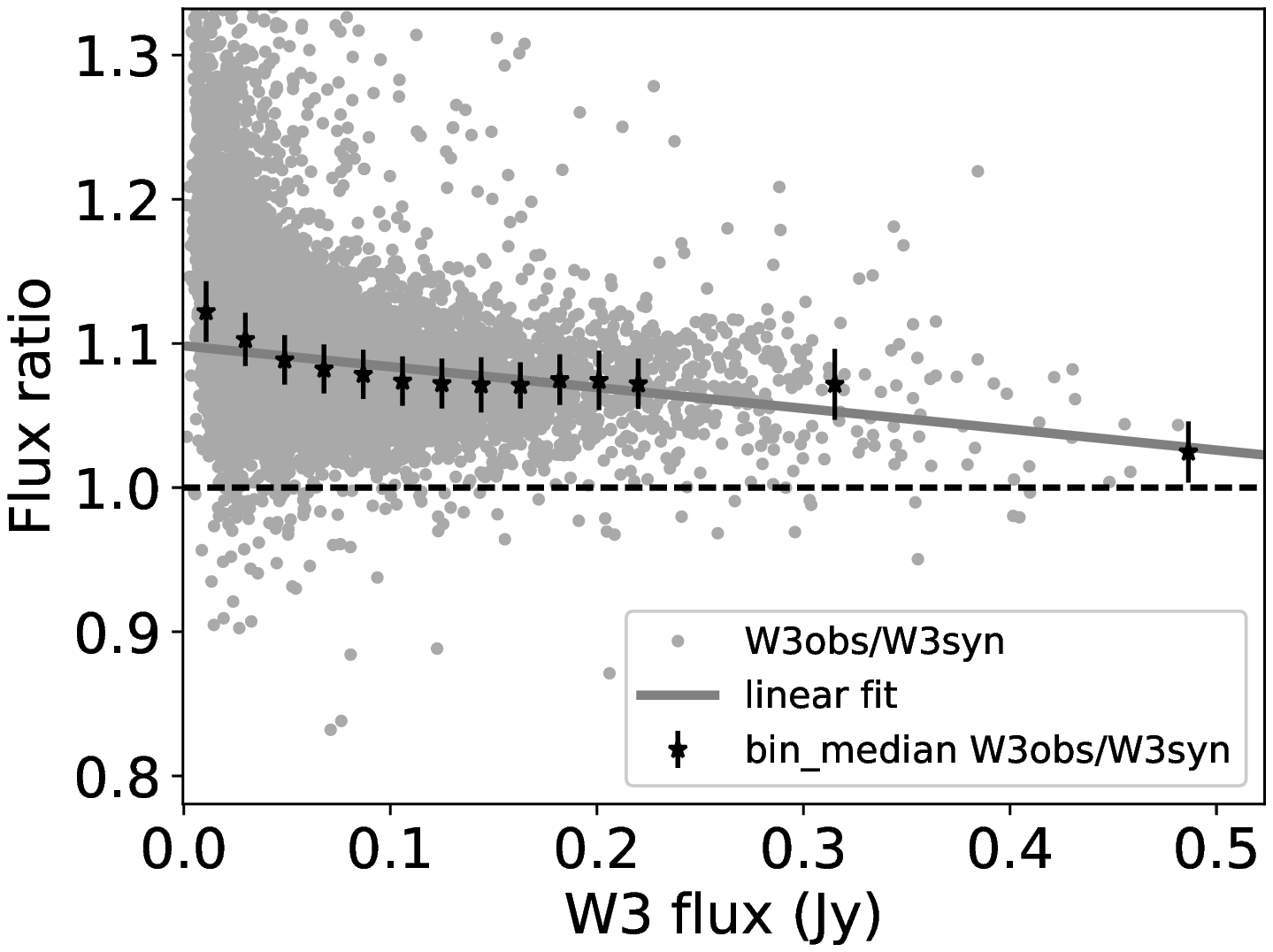} &
\includegraphics[width=9cm]{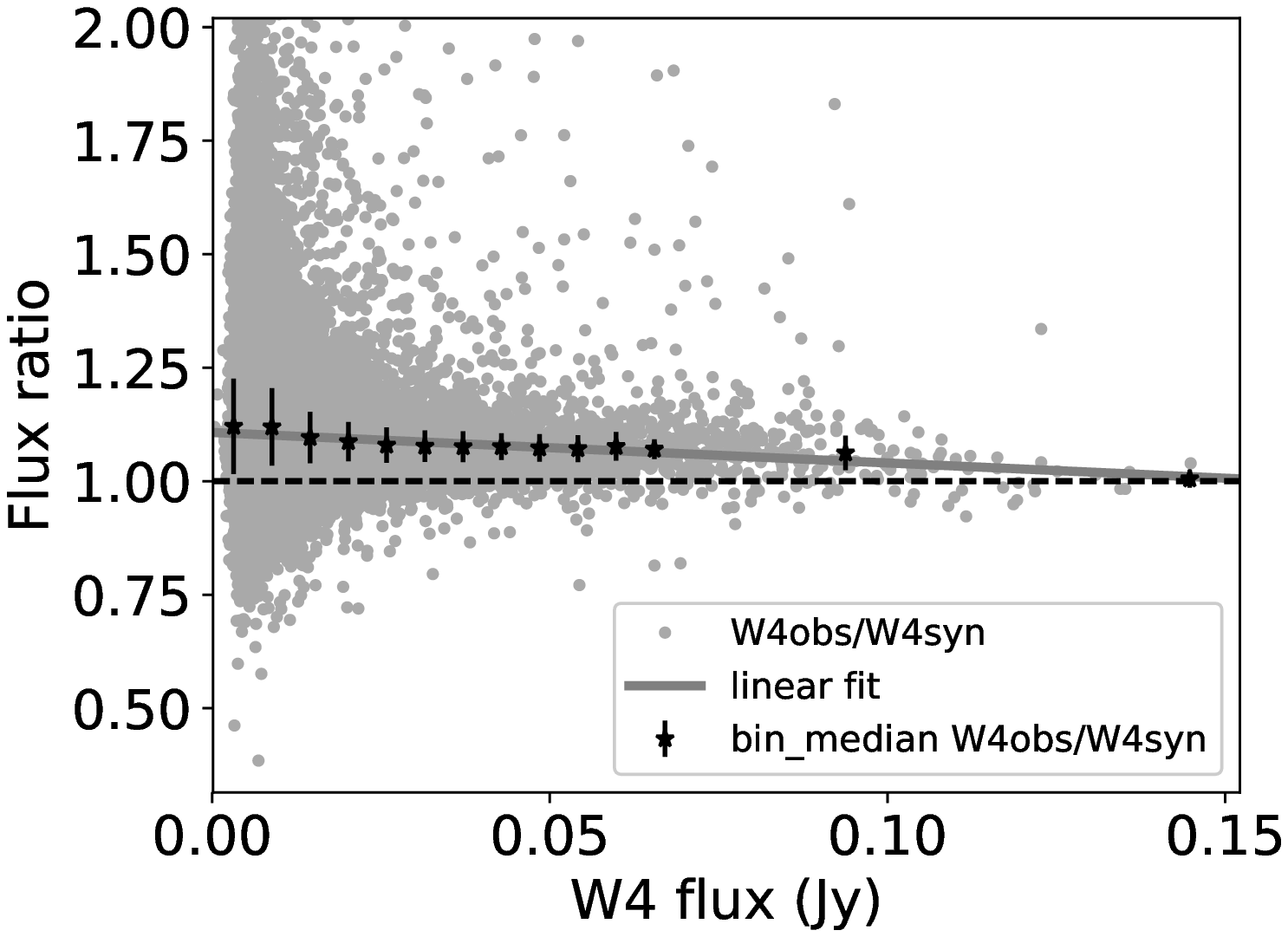} \\
\end{tabular}
\caption{Flux ratio $f_{\rm Wi,obs}/f_{\rm Wi,syn}$ vs. $f_{Wi,obs}$  for all 24,117 stars (dots). The starred symbols represent 
the median for each flux bin with their respective uncertainty, the thick line shows the best linear fit  while the dashed line indicates the reference flux ratio equal to one.}
\label{w2med}
\end{center} 
\end{figure*}

We derive the effective temperature ($T_{\rm eff}$) by identifying the theoretical spectrum that best match the observed 
B, V, J, H, and $K_s$ fluxes\footnote{We extract filter transmission functions from 
http://voservices.net/filter/ and http://www.ipac.caltech.edu/2mass.}, that we assume as purely photospheric, 
using the MPFIT $\chi^2$ minimization algorithm \citep{markwidth2009}. The procedure also provides the multiplying 
factor $c_n$ that takes into account the flux density dependence on distance. Since the surface gravity ($\log{g}$) and 
overall metallicity ([M/H]) generate small flux changes in the Rayleigh-Jeans regime \citep[see][]{cruz2012}, we fixed 
their values to $\log{g}$=4.0 for main sequence (MS) A and F stars and for stars with luminosity class IV or IV-V, and 
$\log{g}$=4.5 for MS G and K stars. We assumed solar metallicity in all cases.

We used the theoretical photospheric SED of each star to compute the {\it WISE} synthetic 
photometry\footnote{http://http://wise2.ipac.caltech.edu/docs/release/allsky/expsup}. In Fig.~\ref{w2med} we plot 
the results in the form of the flux ratio between the observed and the photospheric value as a function of the flux 
value for W1, W2, W3, and W4 bands  for all 24,117 stars. We note a significant discrepancy in all cases. It is smaller for the W1 band and 
it decreases towards brighter objects, apart for the case of W2 band, where it steeply increases with increasing flux. 
To quantify this discrepancy, we divided the sample in flux bins of 0.1~Jy or wider, when the number of objects per 
bin fell short of 40; we then computed the median flux ratio for each bin and we estimated its uncertainty as half the 
difference between the 25$\%$ and 75$\%$ quartiles. We used these data to search for the best linear fit using the MPFIT 
algorithm, with the slope $m$ and the constant $b$ as free parameters in the linear 
equation $f_{\rm i, obs}/f_{\rm Wi,syn}=mf_{\rm i, obs}+b$, where $i$ indicates the {\it WISE} band. The use of the median value 
minimizes the impact of the objects with a real IR excess on the fit, that therefore should be representative of the 
majority of stars whose flux emerges from the photosphere. The results, reported in Table~\ref{param}, show a similar 
behavior of the two redder bands: they have a 10$\%$ flux overestimation over the synthetic value at low fluxes and 
this discrepancy gradually diminishes for the brigther objects. Conversely, for the case of the W2 band, the 
theoretical and observed flux match well for the faint objects, but increases to about a 40$\%$ overestimation 
at the brightest end.

Since, this discrepancy may significantly affect the determination of IR excesses, we decided to transform the {\it WISE} 
observed fluxes to the theoretical system, by means of the following equation:
\begin{equation}
f^\prime_{\rm i, obs} = \frac{ f_{\rm i, obs} }{ mf_{\rm i, obs}+b },
\end{equation}
where $f^\prime_{\rm i, obs}$ is the transformed flux that we use to explore the {\it WISE} database in search of IR excesses 
in stars with and without planets.

\begin{table}
\begin{center}
\caption{Linear fit parameters.}
\label{param}
\begin{tabular}{c c c }
\hline
{\it WISE} band & {\bf $m$} &{\bf $b$} \\
\hline
W1 & -0.008$\pm$0.004 & 1.045$\pm$ 0.009\\
W2 &  0.105$\pm$0.012  & 1.017$\pm$ 0.010 \\
W3 & -0.145$\pm$0.045  & 1.098$\pm$ 0.008 \\
W4 & -0.672$\pm$0.206  & 1.107$\pm$ 0.016 \\
\hline
\end{tabular}
\end{center}
\end{table}

\section{Searching for IR excesses in stars with and without planets }
\label{irno}
 
\subsection{Sample of stars with and without planets}

We constructed the sample of stars hosting planets from ``The Extrasolar Planet Encyclopedia" web page\footnote{http://exoplanet.eu}. 
As of November 2016, the database included 2609 stars with confirmed planets. We applied to this sample the same criteria, on {\it WISE}
and 2MASS photometry, described in Sec.~\ref{selectioncriteria} and we obtained a final sample of 236 objects.  In order to avoid the inclusion of spurious detections, we also proceeded to perform a careful visual inspection of the W4 images of all these stars.
We should mention that 
some well known sources hosting debris discs like Fomalhaut and $\beta$~Pic were not considered in the sample because these 
stars are very bright sources and presented saturated photometry, thus they could not pass the selection criteria defined in section 
\ref{selectioncriteria}.

To build a comparison sample of stars without planets, we considered objects that were included in planet-search projects but 
for which no planetary companions were found. We considered the works by \citet{sousa2008, sousa2011, adibekyan2012, santos2011, 
bertran2015}, that used the High Accuracy Radial velocity Planet Searcher (HARPS) spectrograph, the sample of \citet{valenti2005}, 
that made use of observations with the Keck, the Anglo-Australian and the Lick telescopes, and the study of \citet{wittenmyer2011}, 
who also used the Anglo-Australian telescope as part of the Pan-Pacific Planet Search programme. We eliminated the duplicate entries 
and those stars for which a planet has been discovered after the date of publication. To this initial list of 1850 objects we also 
applied the same selection criteria as before and the sample of stars without planets ended with 986 stars.

\begin{figure*}
\begin{center}
\begin{tabular}{cc}
\includegraphics[width=8cm, height=5.5cm]{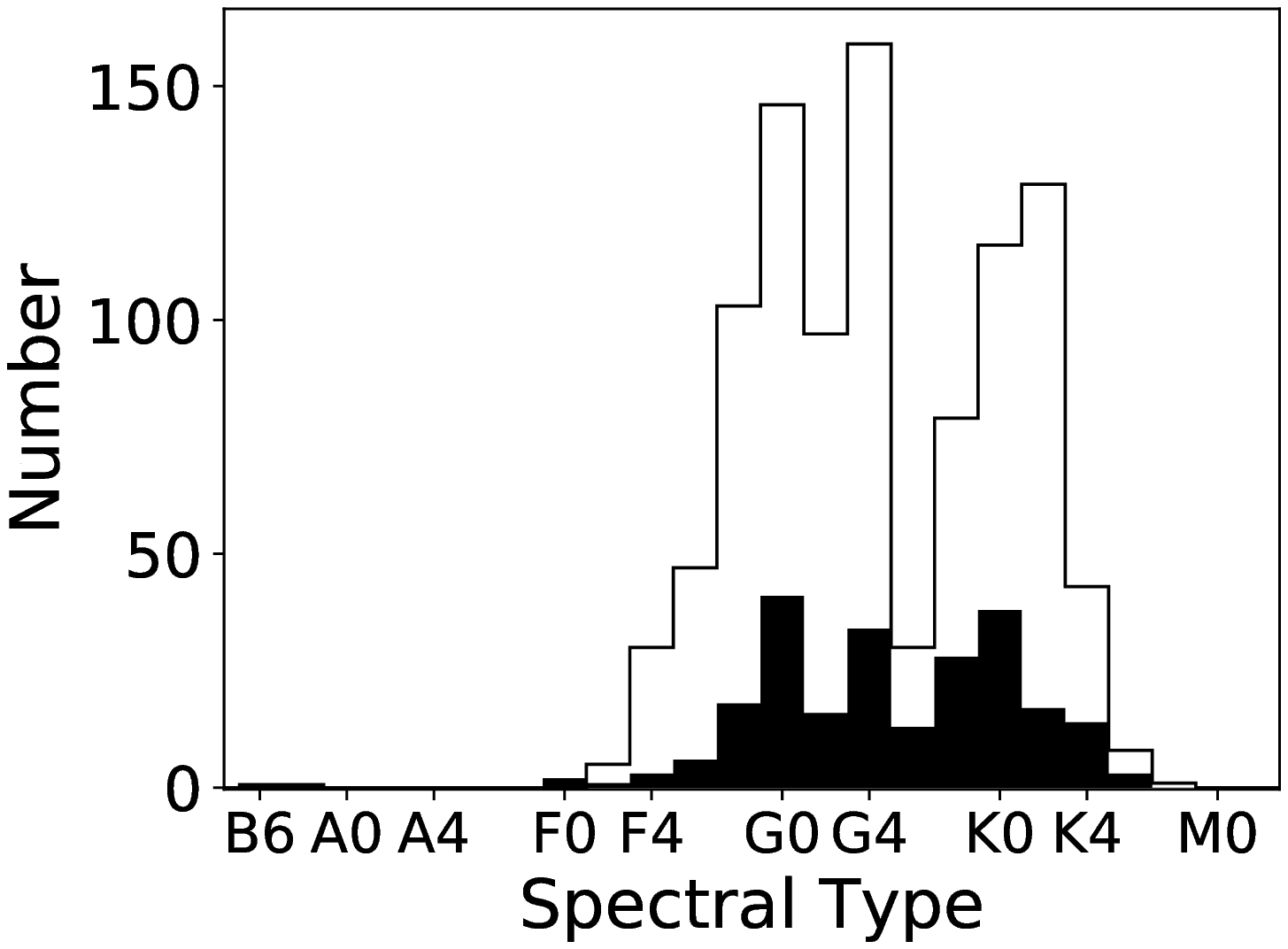} &
\includegraphics[width=8cm, height=5.5cm]{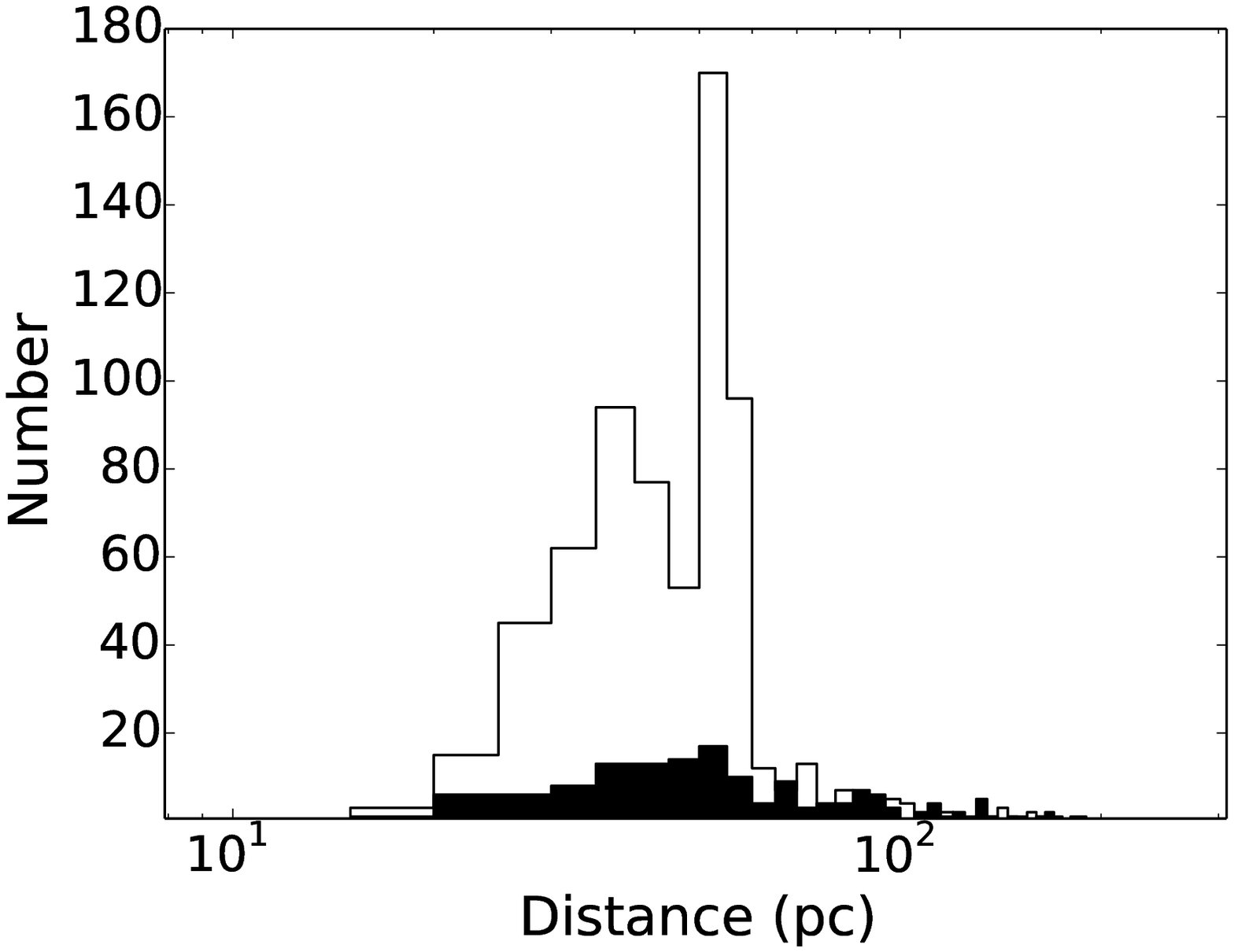} \\
\includegraphics[width=8cm, height=5.5cm]{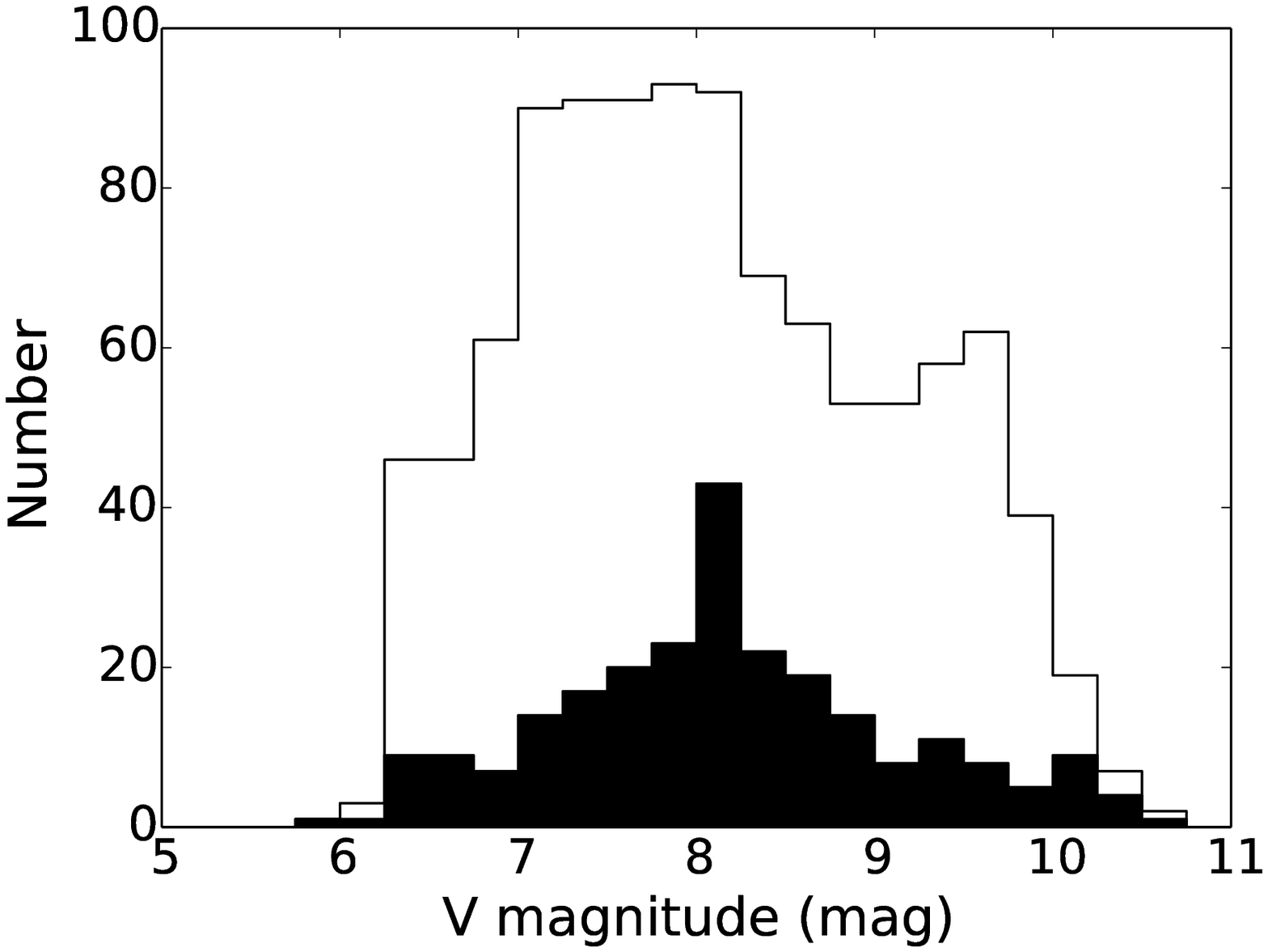} & 
\includegraphics[width=8cm, height=5.5cm]{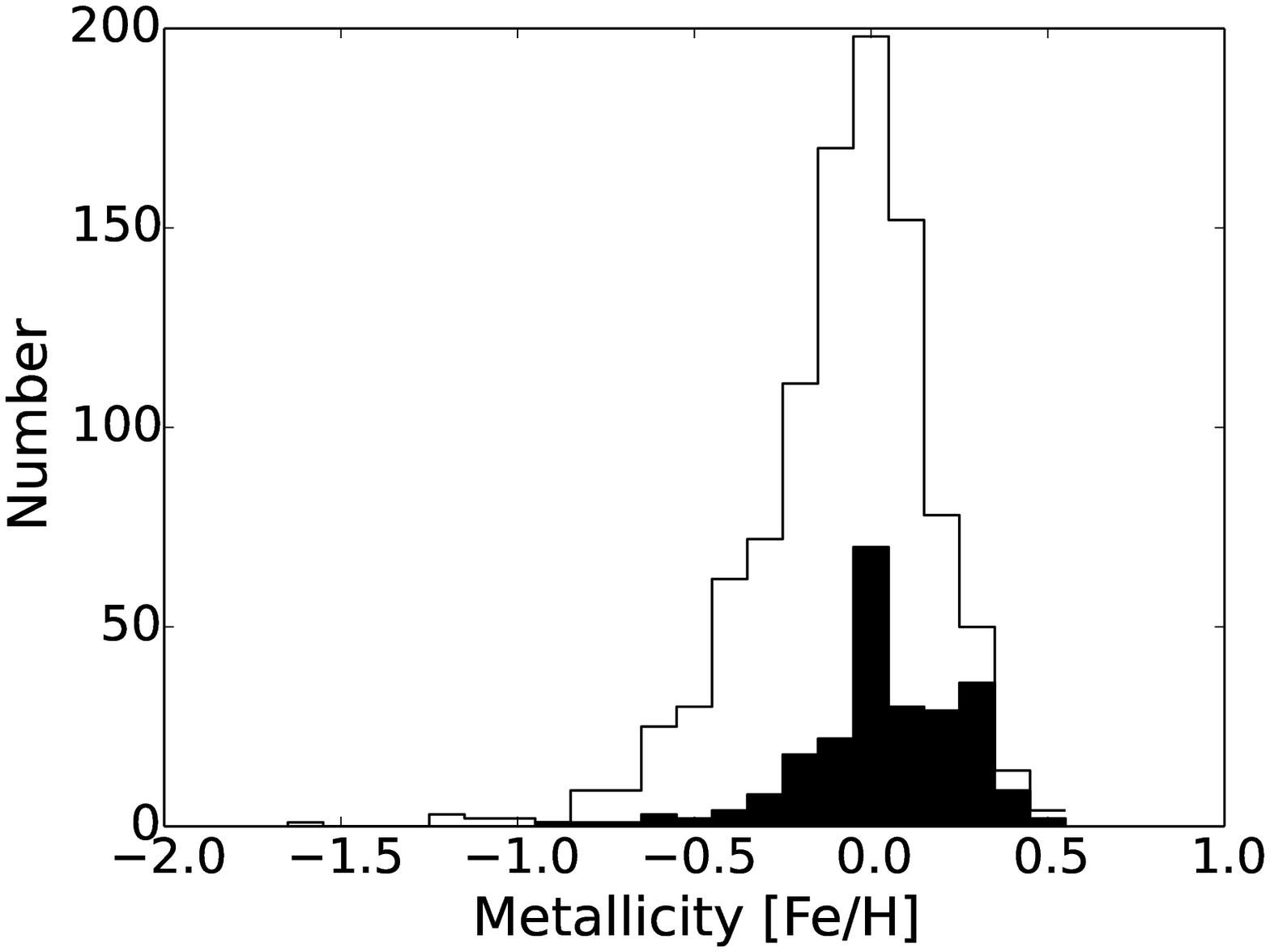} \\
\end{tabular}
\caption{Distributions of the 236 stars with planets and of the 986 stars without planets with respect to spectral type, distance, apparent V magnitude and metallicity.}
\label{fig24}
\end{center} 
\end{figure*}

In Fig.~\ref{fig24} we plot the distribution of the two samples with respect to spectral type, distance, apparent V magnitude and 
metallicity ([Fe/H]). We applied a two-sample Kolmogorov Smirnov (KS) test to evaluate whether the samples of stars with and 
without planets share the same parent population, considering a confidence level threshold of 99$\%$ \citep{massey1952}. This 
hypothesis was verified in the case of the spectral type and V magnitude distributions, while discarded in the distance and 
metallicity cases. As far as the distance is concerned, the distribution of Fig.~\ref{fig24} shows that the sample of stars 
without planets is composed on average by more nearby stars with a mean of 50~pc while the other sample has a mean distance of 
75~pc. The cut at the larger distance is due to the photometry criteria that we imposed. In the case of the metallicity, the 
discrepancy between the two samples is more evident: the average metallicity of the stars without planets is [Fe/H]=$-$0.09, 
in agreement with the typical value in the solar vicinity \citep[e.g.,][]{casagrande2011,haywood2001}, while the stars with 
planets are more metal-rich, with an average  [Fe/H]=+0.04. This results concords with the positive correlation that exists 
between the presence of giant planets and the metal content of the host stars \citep[e.g.,][]{gonzalez1997,valefisch2005}. 
Since about 60$\%$ of stars in our sample have planets with mass $\geq 1$~M$_J$, we expected to find this metal abundance 
difference in our samples.

\subsection{Looking for IR excesses.}

In this section, we describe the method we devised for identifying stars with IR excess from the All{\it WISE} data. To date, 
there have been several works that used {\it WISE} data with the same goal. They differ from our work and among them in the 
definition of IR excess. We therefore considered convenient to also incorporate the procedures used by \citet{cruz2014}, 
\citet{kennedy2012}, \citet{morales2012} and \citet{patel2014} in order to understand how their differences affect the 
number of detected IR excesses.

\subsubsection{Excess definition 1: this work.}

We define a IR excess value $E$ as the difference of the $f_{22}$/$f_{12}$ observed flux ratio and the respective stellar 
photospheric ratio, divided by the associated uncertainty:
\begin{equation}
E = \frac{\bigl(\frac{f_{22}}{f_{12}} \bigr)_{obs}-\bigl(\frac{f_{22}}{f_{12}}\bigr)_{syn}}{\sigma_{tot}},\quad  \sigma_{tot}=\sqrt{\sigma_{obs}^2+\sigma_{cal}^2+\sigma_{syn}^2},
\end{equation}
where $(\frac{f_{22}}{f_{12}})_{obs}$ is the observed flux ratio  between W4 and W3 bands, $(\frac{f_{22}}{f_{12}})_{syn}$ is the synthetic flux ratio. 
$\sigma_{tot}$ is the quadratic sum of 3 error sources: 1) the  photometric contribution $\sigma_{obs}$, obtained 
from the {\it WISE} catalogue; 2) the calibration uncertainty of {\it WISE} bands $\sigma_{cal}$ of 4.5$\%$ and 5.7$\%$ for 
W3 and W4, respectively \citep{jarret2011}; 3) the uncertainty due to the stellar parameters $\sigma_{syn}$. 
This latter value is obtained by first constructing a family of synthetic spectra, whose parameters were a 
combination of $T_{\rm eff}\pm\sigma_{\rm Teff}$ and $c_n\pm\sigma_{c_n}$ (see Sec.~\ref{photfitting}). Then, 
the synthetic flux is calculated for each spectrum and the standard deviation of the flux distribution is 
considered as $\sigma_{syn}$. An infrared excess is considered significant when $E$ is larger than a chosen threshold. 

We adopted the $f_{22}$/$f_{12}$ ratio because these 2 bands show a quite similar trend in the flux ratio between observed and 
photospheric flux (see Fig.~\ref{w2med}). However, this choice may affect the detection of the hotter discs, if 
they emit a non negligible amount of flux at 12~$\mu$m.

\subsubsection{Excess definition 2: \citet{cruz2014}.}

The authors searched for IR excesses in a sample of dwarf stars brighter than V=15~mag, using the 
{\it WISE} All-Sky survey database for over 9000 sources. They compared the observed $f_{22}$/$f_{4.6}$ flux ratio with the 
expected photospheric level and identified 197 significant excesses at 22~$\mu$m. Assuming that the excess 
is caused by thermal emission of a circumstellar dust disc, they found more than 80$\%$ of the sample with 
dust temperatures higher than 120~K. The definitions of flux excess and its error used by \citet{cruz2014} 
is the same as those adopted in this work, with the only difference that the flux ratio they considered is $f_{22}$/$f_{4.6}$.

\subsubsection{Excess definition 3: \citet{kennedy2012}.}

They analyzed 180\,000 stars observed by {\it WISE} in the Kepler field. After carefully taking 
into account contamination by background sources, they identify only one bonafide excess at 22~$\mu$m due 
to a debris disc around the A-type star KIC\,7345479. They defined an excess as the difference of the observed 
flux in any {\it WISE} band and the correspondent photospheric flux, divided by the total error:
\begin{equation}
E=\frac{f_{i,obs}-f_{i,syn}}{\sigma_{tot}}
\end{equation}
where $\sigma_{tot}$ is the quadratic sum of the observed photometric uncertainty $\sigma_{obs}$ and the 
uncertainty of the expected stellar flux, equivalent to the error produced by the uncertainty on stellar 
parameters $\sigma_{syn}$. They adopted a $4\sigma$ threshold.

\subsubsection{Excess definition 4: \citep{morales2012}.}

This work explored exoplanet-host stars in sear\-ching for the presence of warm dust. 
They analyzed a sample of 591 stars from the Extrasolar Planet Encyclopedia (January 2012) and found 
9 planet-bearing stars with excesses at 12~$\mu$m and 22~$\mu$m in young MS and giant stars. 

To detect IR excess, after obtaining the synthetic spectrum that better fit the photosphere, they 
calculated the significance of the excess using the total signal-to-noise  ratio (S/N) defined as:
\begin{equation}
\biggl(\frac{S}{N}\biggr)=\frac{f_{obs}}{\sigma}=\biggl(\frac{S}{N}\biggr)_{syn}+\biggl(\frac{S}{N}\biggr)_d,\quad \quad \quad \sigma=\sqrt{\sigma_{obs}^2+\sigma_{cal}^2} \, ,
\end{equation}
with $f_{obs}=f_{syn}+f_d$, where $f_{syn}$ is the stellar synthetic flux of the best fit, $W_d$ is the flux of a 
debris disc in any of the {\it WISE} bands, and $\sigma$ takes into account the uncertainty of the {\it WISE} photometry 
($\sigma_{obs}$) and calibration ($\sigma_{cal}$). 
The authors considered an infrared excess when $(S/N)_d\geq 3$.

\subsubsection{Excess definition 5: \citep{patel2014}.}

The last work we analyzed is \citet{patel2014},  where they searched for excesses, in the {\it WISE} database at 12~$\mu$m and 22~$\mu$m, 
of Hipparcos MS stars within 75~pc from the Sun. Since they were dealing with many bright objects, they apply 
corrections for properly handling with saturated objects in  W1 and W2 bands. They found 379 stars with flux excesses at 22~$\mu$m. 

The procedure they devised for detecting infrared excesses is quite different than the previous ones. 
\citet{patel2014} based their searching process on empirical relations found in colour-colour diagrams of 
Tycho-2 $B-V$  vs. $(W_i-W_j)$, with $i,j$ representing two different {\it WISE} bands. They did not carry out a 
fitting on the star SED like the previous methods to determine the photospheric flux level, however, they 
fit the star SED to confirm the validity of the detected excesses.

The authors have calibrated  the relations $W_i-W_j$ vs. $B-V$ by iteratively removing  the largest colours 
$W_i-W_j$, for each  $B-V$ bin of 0.1~mag, until half of the data points in the bin are rejected. The different 
relations $W_i-W_j$ vs. $B-V$ are traced in steps of 0.02~mag in $B-V$. Thus, the mean $W_i-W_j$ corresponding 
to a certain $B-V$ colour is referred as  $W_{ij}(B-V)$. The list of the mean $W_{ij}(B-V)$, with its respective 
standard error, for all colour combinations is given also in \citet{patel2014}. Taking into account the colour 
dependency, an excess $E[W_i-W_j]$ in the colour  $W_i-W_j$ for  a certain $B-V$ is defined as:
\begin{equation}
E[W_i-W_j]=W_i-W_j-W_{ij}(B-V) \, , 
\end{equation}
while, the S/N  of this excess is:
\begin{equation}
\frac{S}{N}=\frac{E[W_i-W_j]}{\sigma_{ij}}=\frac{W_i-W_j-W_{ij}(B-V)}{\sigma_{ij}}
\end{equation}
where $\sigma_{ij}$ is the propagation of the photometric uncertainties of $W_i$ and $W_j$, 
together with the standard error $W_{ij}(B-V)$; thus,
\begin{equation}
\sigma_{ij}=\sqrt{\sigma_{Wi}^2+\sigma_{Wj}^2+\sigma_{Wij}^2} \, .
\end{equation}

They calculated the ratio of detecting false positives (FPR) by using the empirical S/N distribution and defined FPR  as the number of stars  beyond the threshold in  which the outliers were considered as reliable excesses. 
This enabled to detect excesses as a function of the threshold beyond which the redder objects can be considered 
as reliable excesses. Nevertheless, it was not possible to determine empirically the FPR beyond the limit where 
the number of false positives drops to zero; for this, an upper limit was needed. To define this limit, the 
distributions of S/N, which involved W4 as $W_j$, were constructed and the threshold was located between 
99.8$\%$ and 99.9$\%$ for considering a possible excess.

\section{Results}

We applied all five procedures described in the previous section to the search for IR excess in our 
two stellar samples of stars with and without planets, using the All{\it WISE} catalogue \citep{cutri2013}. 
We present the results in Table~\ref{tab41}; for sake of homogeneity, we used the same threshold 
for all methods and we assumed both $3\sigma$ and $4\sigma$ as detection limits. We report the 
number and percentage of significant IR excesses detected using both the 
modified {\it WISE} flux (as explained in Sec.~\ref{sec:fluxcorr}), in columns 2--3 and 6--7, and the original All{\it WISE} 
photometry, in columns 4--5 and 8--9.

\begin{table*}
\begin{center}
\caption[Number and percentage of IR excesses in stars with and without planets.]{\footnotesize Number and percentage of 
IR excesses in stars with and without planets, using 3$\sigma$ (top) and 4$\sigma$ (bottom) threshold.}
\label{tab41}
\begin{tabular}{l r@{}l c c c  r@{}l c c c}
\noalign{\smallskip} \hline \noalign{\smallskip}
{ IR excess definition} &\multicolumn{5}{|c|}{ Stars with planets}&\multicolumn{5}{|c|}{ Stars without planets}\\
 \hline \noalign{\smallskip}
   & \multicolumn{3}{|c|}{corrected flux} & \multicolumn{2}{|c|}{All{\it WISE} flux} & \multicolumn{3}{|c|}{corrected flux} & \multicolumn{2}{|c|}{All{\it WISE} flux} \\
\hline \noalign{\smallskip}
& \multicolumn{2}{|c|}{\#} &  $\%$ & \# & $\%$ & \multicolumn{2}{|c|}{\#} & $\%$ & \# & $\%$ \\
 \hline \noalign{\smallskip}
\multicolumn{11}{|c|}{3$\sigma$ threshold}\\
 \hline \noalign{\smallskip}
This work                       & 2&$^{a,b}$   & 0.85 & 2 & 0.85 & 1&$^d$           & 0.10 & 1  & 0.10 \\
Cruz-Saenz de Miera et al. 2014 & 1&$^a$       & 0.42 & 1 & 0.42 & 1&$^e$           & 0.10 & 1  & 0.10 \\ 
Kennedy $\&$ Wyatt 2012         & 2&$^{a,b}$   & 0.85 & 6 & 2.54 & 4&$^{d,e,f,g}$   & 0.41 & 35 & 3.55  \\
Morales et al. 2012             & 2&$^{a,b}$   & 0.85 & 2 & 0.85 & 3&$^{d,e,f}$     & 0.30 & 5  & 0.51 \\
Patel et al. 2014               & 3&$^{a,b,c}$ & 1.27 & 3 & 1.27 & 5&$^{d,e,f,g,h}$ & 0.51 & 5  & 0.51  \\
\noalign{\smallskip} \hline \noalign{\smallskip}
\multicolumn{11}{|c|}{4$\sigma$ threshold}\\
 \hline \noalign{\smallskip}
This work                       & 1&$^a$     & 0.42 & 1 & 0.42 & 0&             & 0    & 0 & 0 \\
Cruz-Saenz de Miera et al. 2014 & 1&$^a$     & 0.42 & 1 & 0.42 & 0&             & 0    & 0 & 0 \\
Kennedy $\&$ Wyatt 2012         & 2&$^{a,b}$ & 0.85 & 2 & 0.85 & 4&$^{d,e,f,g}$ & 0.41 & 9 & 0.91 \\
Morales et al. 2012             & 2&$^{a,b}$ & 0.42 & 2 & 0.85 & 0&             & 0    & 3 & 0.30 \\
Patel et al. 2014               & 2&$^{a,b}$ & 0.85 & 2 & 0.85 & 4&$^{d,e,f,g}$ & 0.41 & 4 & 0.41 \\
\noalign{\smallskip} \hline
\multicolumn{11}{|l|}{$^a$HD106906, $^b$HD218396, $^c$BD-10\,3166, $^d$HD85301, $^e$HD136544, $^f$HD107146, $^g$HD60491, $^h$HD11938} \end{tabular}
\end{center}
\end{table*}

The flux homogeneisation between observed and theoretical fluxes (Sec.~\ref{sec:fluxcorr}) decreases the number 
of detected IR excesses  in those cases where the excess is defined through a difference between observed and 
photospheric value, as in \citet{kennedy2012}  and \citet{morales2012}.
It has a lower effect on the other methods, based on colours, and, even though the significance of the IR excesses 
changes, it is not enough to modify the number of objects whose excess surpasses the detection threshold.

In Table~\ref{tab41}, we also report the identification of the stars with IR excess detected with each definition. 
 The one adopted in this work allowed to detect 2 excesses  over the photosphere at 22~$\mu$m in stars with planets (0.85$\%$ of the sample) and just 1 excess (0.1$\%$) in stars that do not harbour a detected planet, assuming a 3$\sigma$ threshold. The excess definition of \citet{cruz2014} provides even lower numbers,  while 
the results from the  method by \citet{patel2014} include all objects detected with the other definitions  (with a detection rate of 1.27$\%$ and 0.51$\%$ in stars with and without planets, respectively) and adds 
two unique stars (BD-10\,3166 and HD11938), if the lower significance threshold of 3$\sigma$ is considered.

\begin{figure*}
\begin{center}
\begin{tabular}{cc}
\includegraphics[width=15cm, height=10cm]{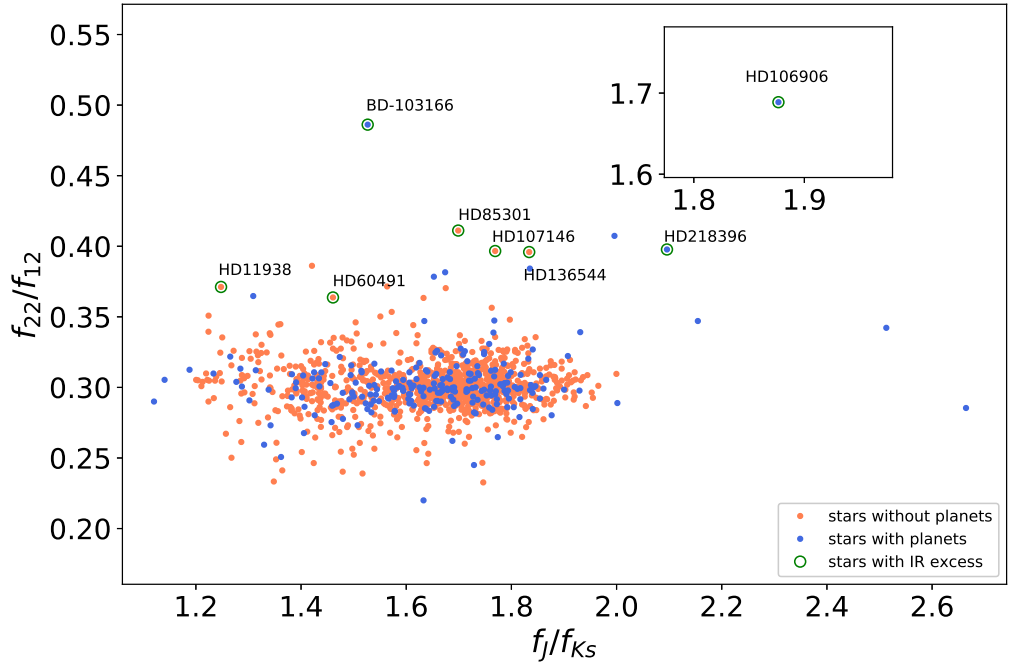} 
\end{tabular}
\caption{Flux ratio diagram of {\it WISE} photometry at 22~$\mu$m and 12~$\mu$m versus
the 2MASS J and $K_s$ fluxes for the 1,222 stars with and without planets, as explained in the inset. Objects with IR excess refer to the stars reported in Table 3. The location of HD 106906 is indicated in the upper inset.}
\label{figcolor}
\end{center} 
\end{figure*}
 
  In Fig.~\ref{figcolor} , we show the flux ratio diagram using the 2MASS J and K$_s$ fluxes for the 1,222 stars with and without planets used in this work and the {\it WISE} fluxes at 12~$\mu$m and 22~$\mu$m.  We highlight the 8 stars with Mid-IR excess reported in Table~\ref{tab41} with circles.

\begin{figure*}
\begin{tabular}{ccc}
\includegraphics[width=5.7cm, height=4.4cm]{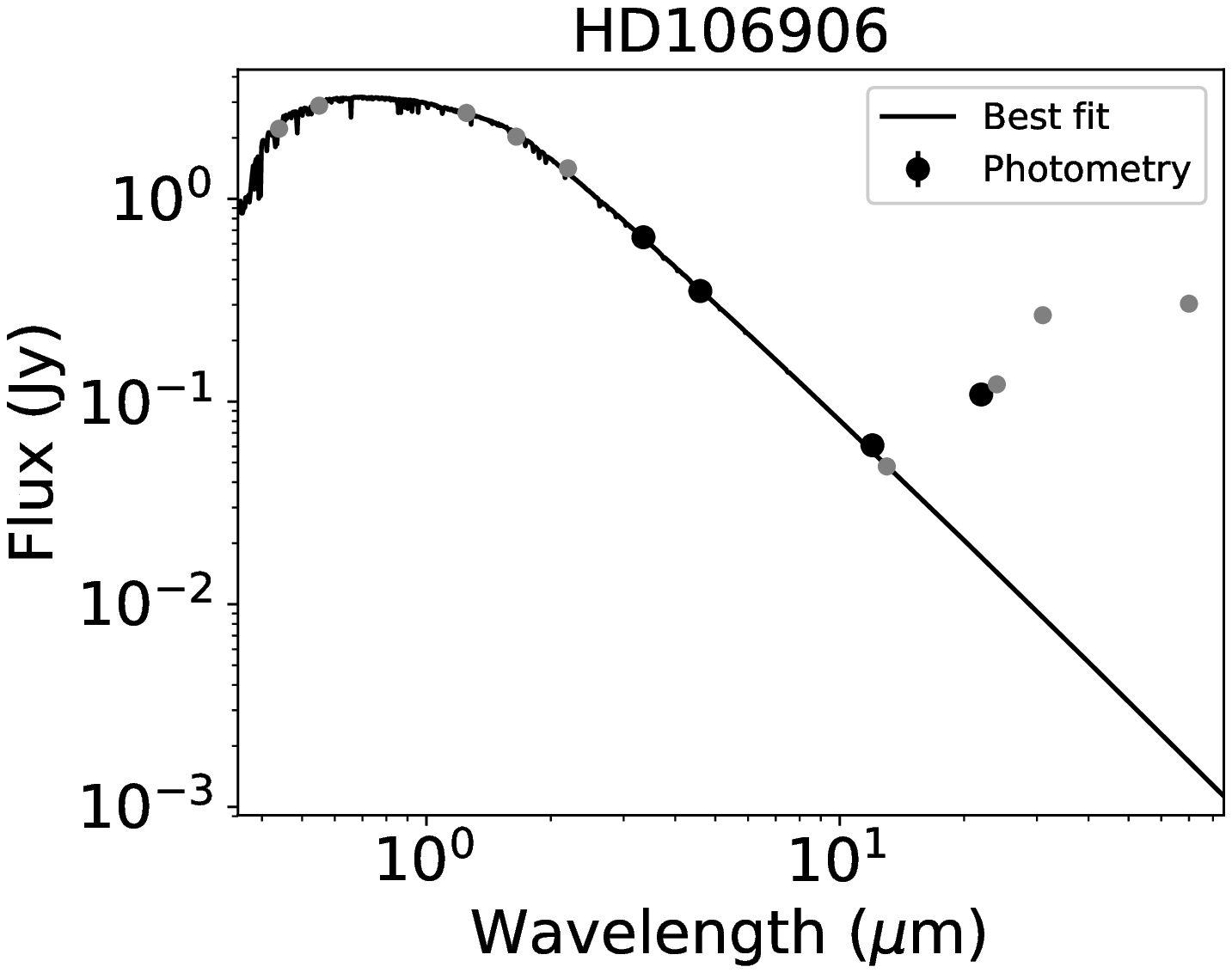} &
\includegraphics[width=5.7cm, height=4.4cm]{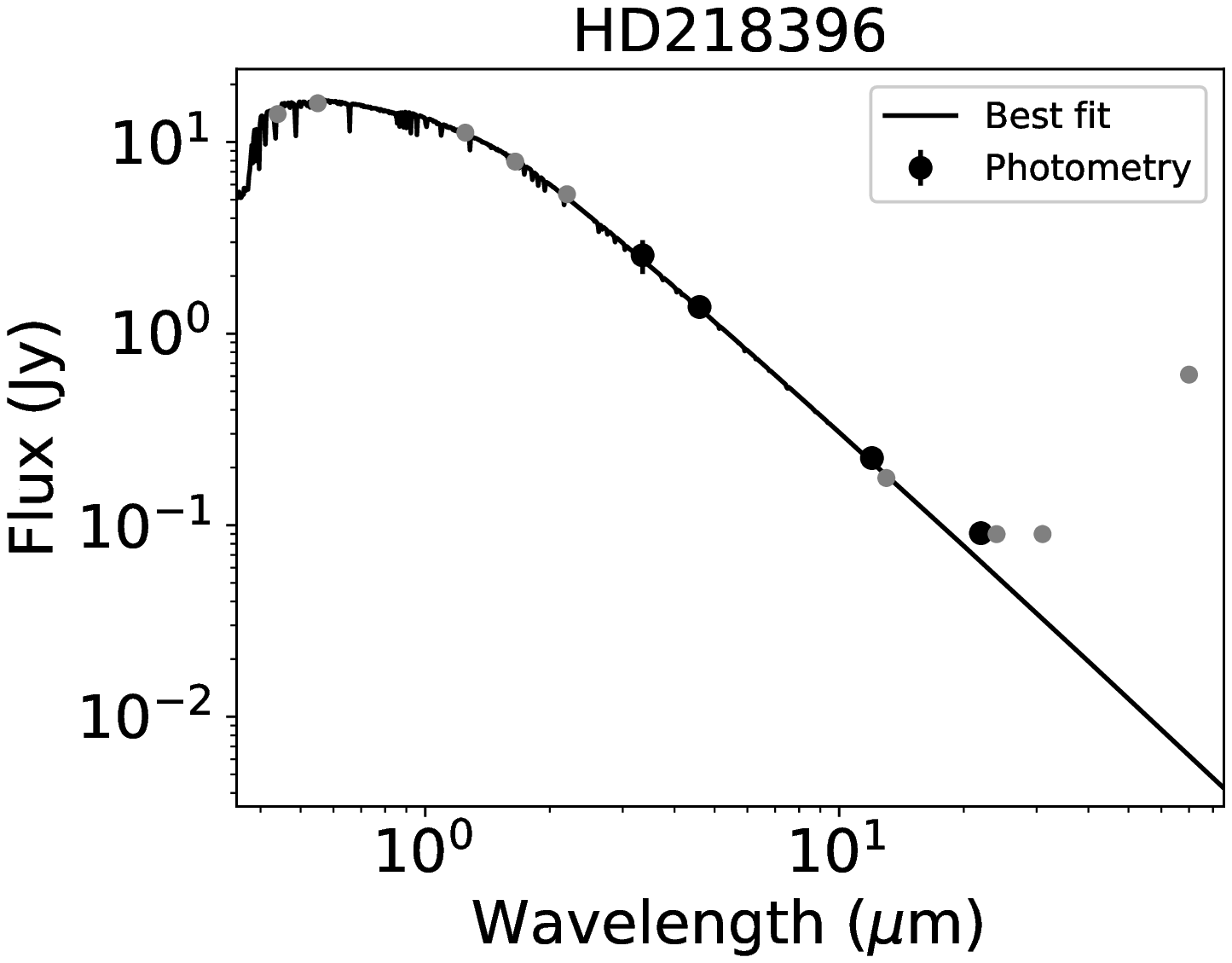} &
\includegraphics[width=5.7cm,height=4.4cm]{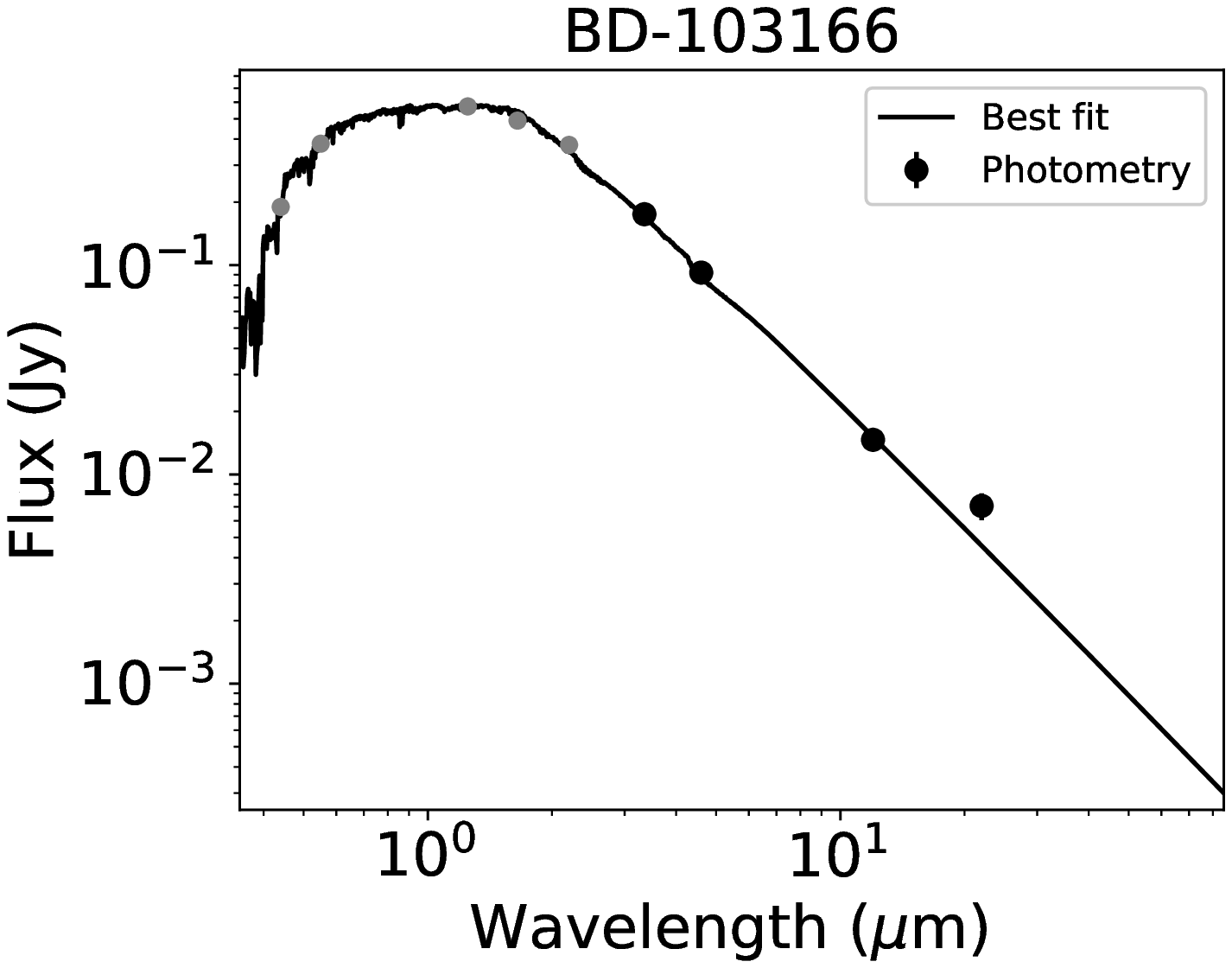} \\
\end{tabular}
\caption{SEDs of the stars with planets for which a significant flux excess has been detected in the {\it WISE} data.  
HD106906, HD218396, BD-10\,3166. We show in black dots the {\it WISE} photometry  while in grey dots optical and 
{\it Spitzer} photometry. Continuous line represents the best atmospheric model fit. }
\label{fig:sed1}
\end{figure*}

\begin{figure*}
\begin{tabular}{ccc}
\includegraphics[width=5.7cm, height=4.4cm]{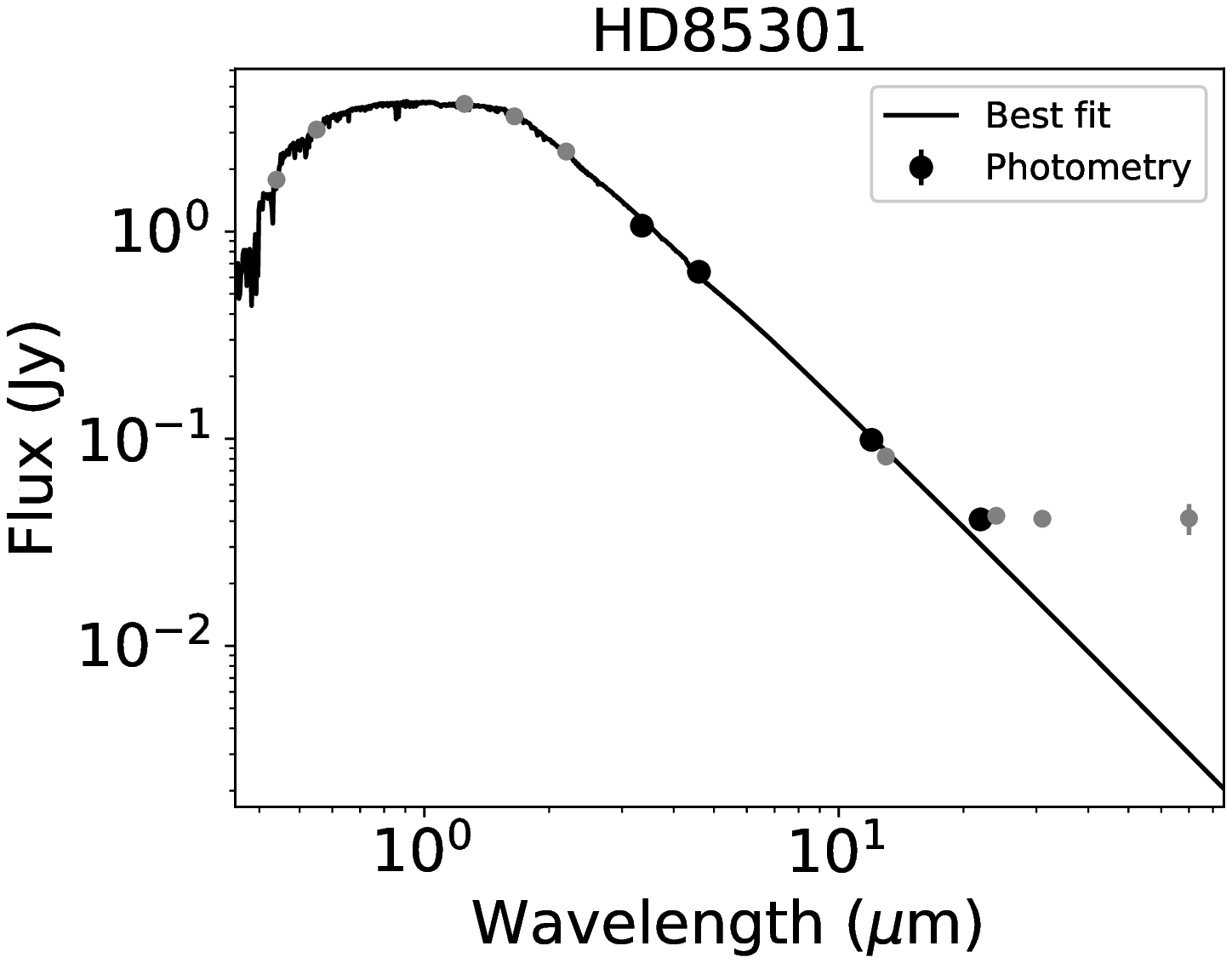} &
\includegraphics[width=5.7cm, height=4.4cm]{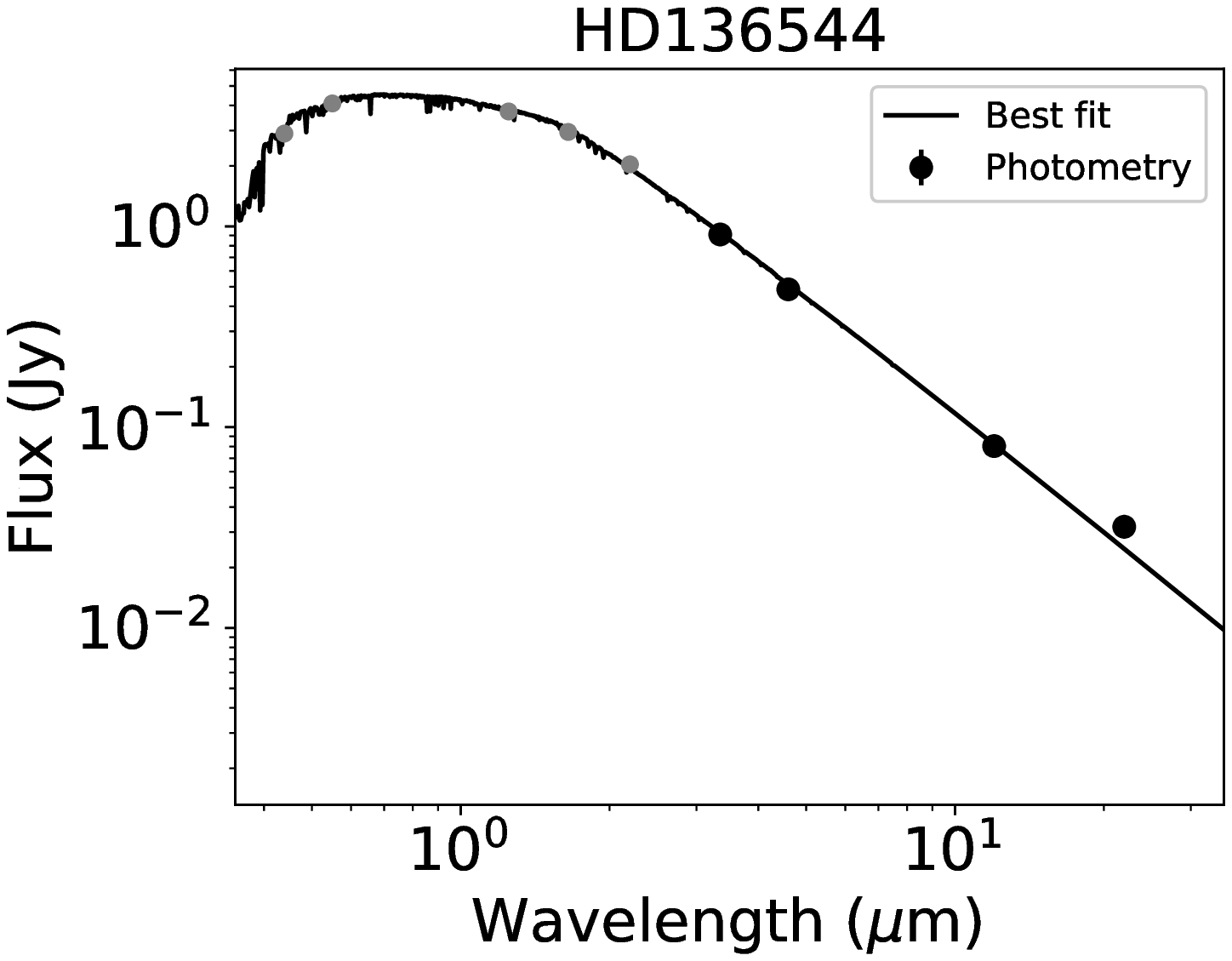} &
\includegraphics[width=5.7cm, height=4.4cm]{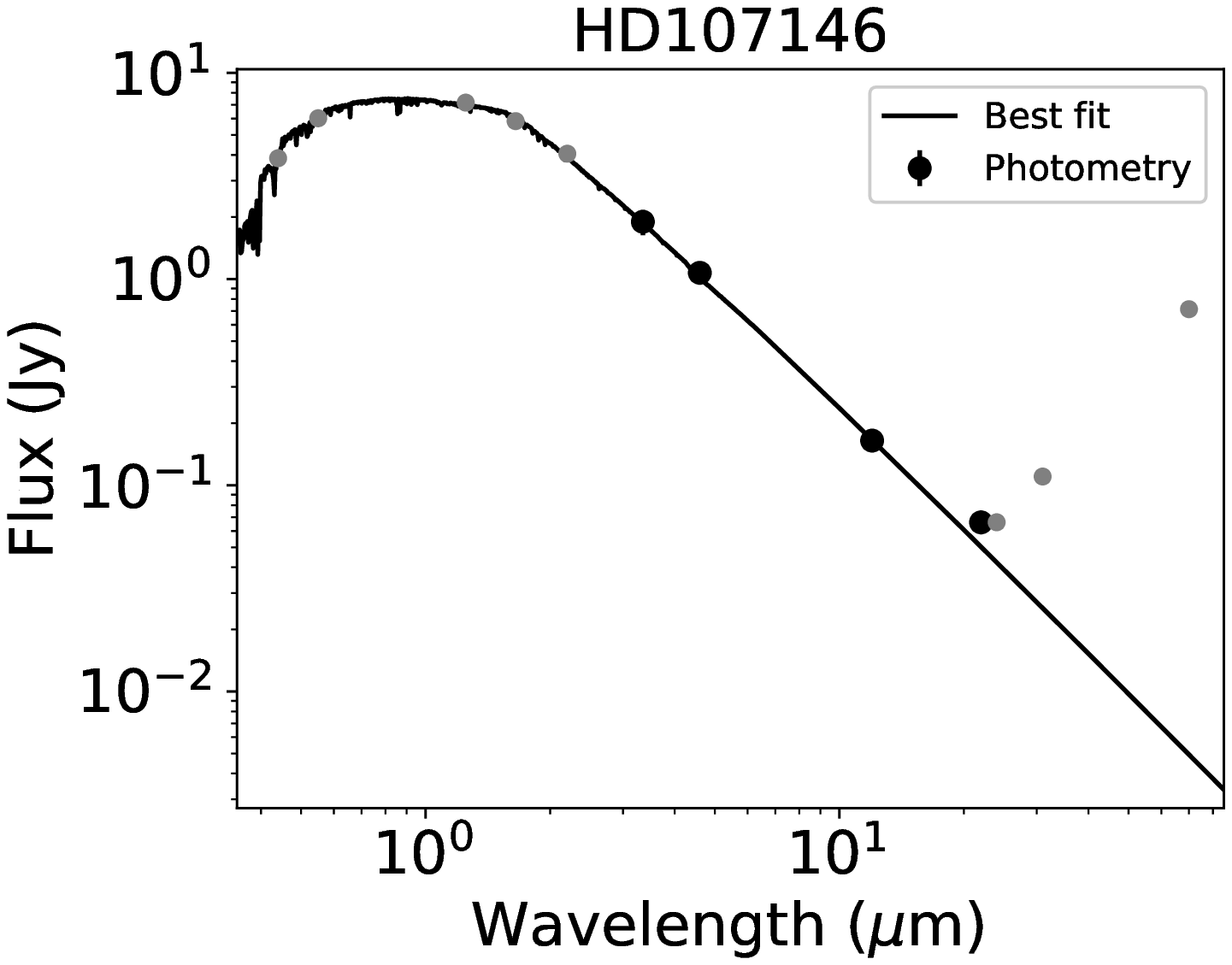} \\
\includegraphics[width=5.7cm, height=4.4cm]{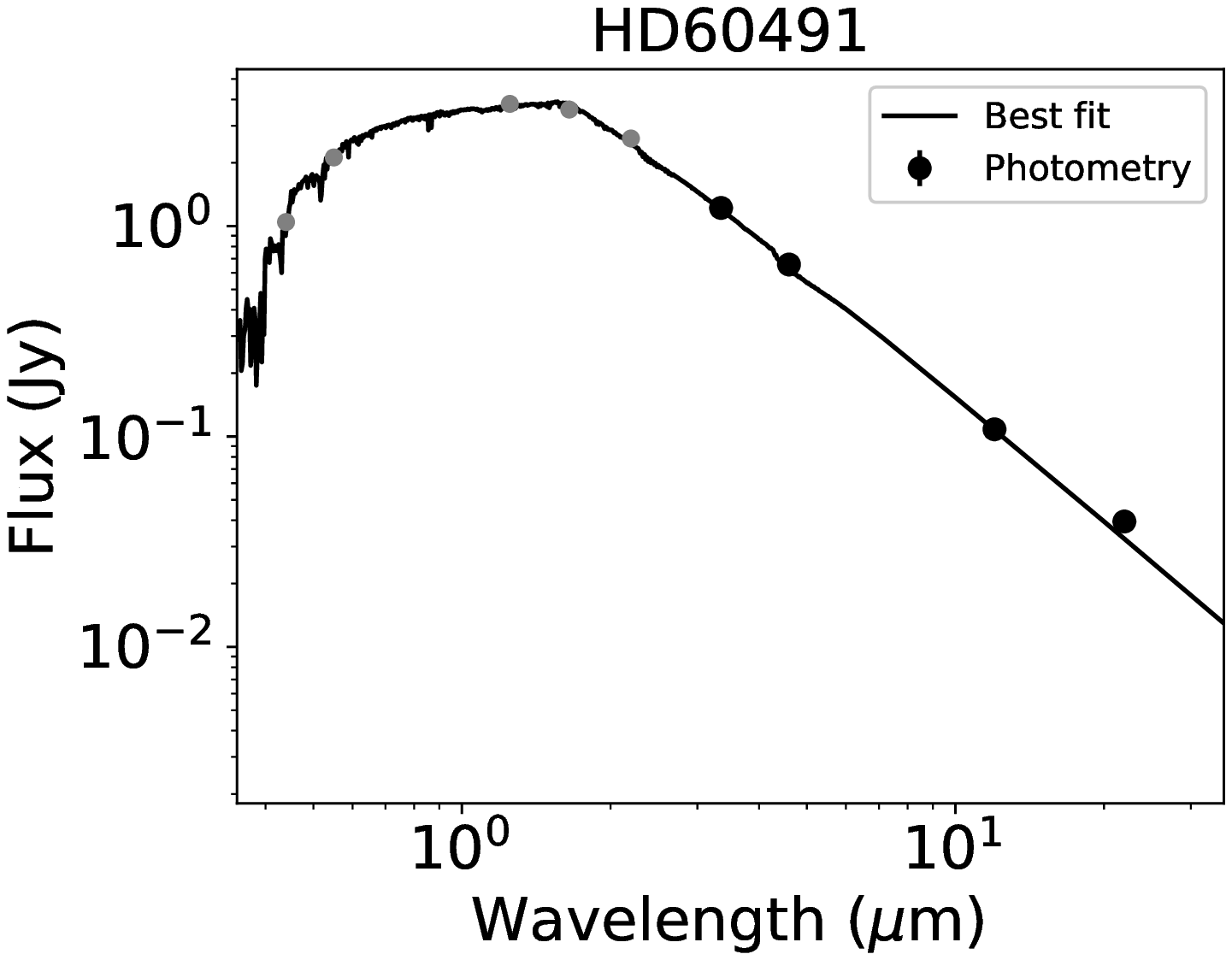} &
\includegraphics[width=5.7cm, height=4.4cm]{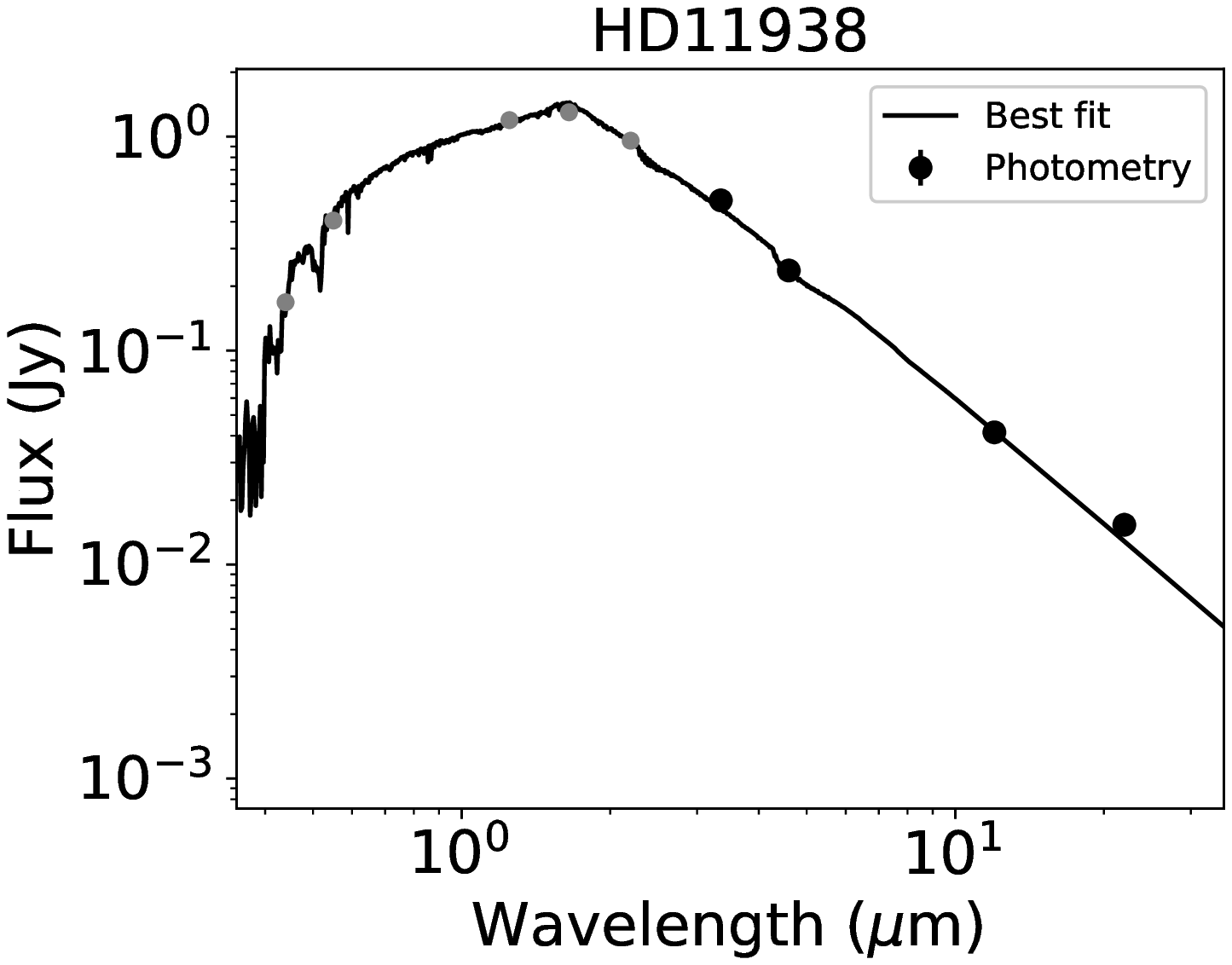} \\
\end{tabular}
\caption{SEDs of the stars without planets for which a significant flux excess has been detected in the {\it WISE} 
data. HD85301, HD136544, HD107146, HD 60491, HD11938. We show in black dots the {\it WISE} photometry  while in grey 
dots optical and {\it Spitzer} photometry. Continuous line represents the best atmospheric model fit.}
\label{fig:sed2}
\end{figure*}

The SED of all 8 objects  found by the method of \citet{patel2014}, using a 3$\sigma$ threshold, are depicted in Figs.~\ref{fig:sed1} and \ref{fig:sed2}.
{\em Spitzer} photometry \citep{chen2005} confirms the presence of an IR excess for 5 stars 
[$^a$HD106906, $^b$HD218396 (HR8799),  $^d$HD85301, $^f$HD107146, $^g$HD60491], with a clear indication that the flux 
peak is located at wavelengths longer than 22~$\mu$m. In fact, apart from the case of HD106906 and HD60491, 
{\em Spitzer} data indicate that the W4 band coincides with the wavelength range where the IR excess  begins to appear above the photosphere, with a $>$3$\sigma$ significance, according to the blackbody modeling by \citet{chen2014}.

\begin{figure}
\begin{center}
\includegraphics[width=9cm]{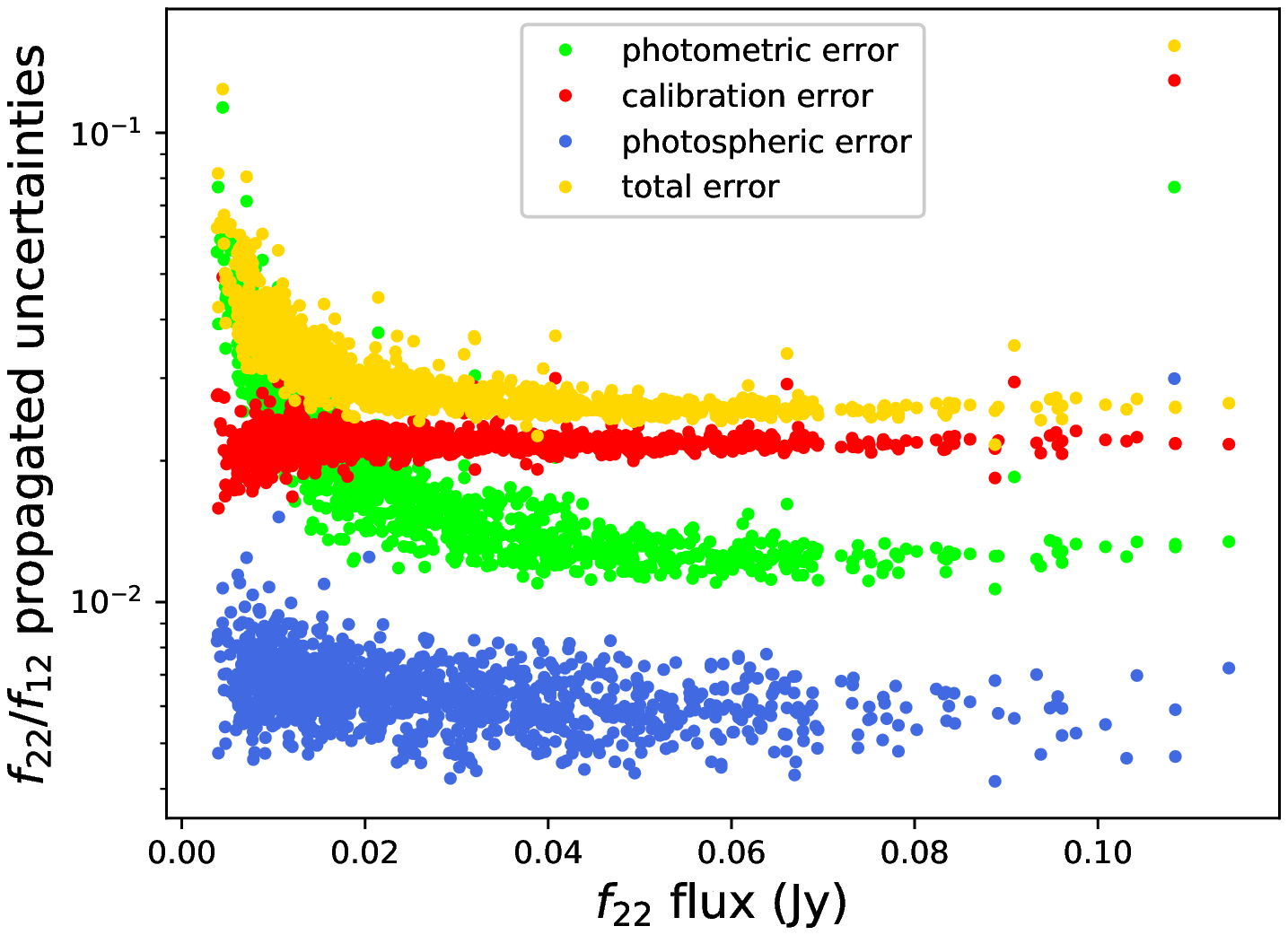}
\caption{Sources of uncertainty in the flux ratio $f_{22}$/$f_{12}$ as a function of the flux  at 22~$\mu$m for the complete sample of 1,222 stars with and without planets. The errors due to {\it WISE} 
photometry (green dots), flux calibration (red) and uncertainty on the stellar photospheric parameters (blue) 
are shown separately, along with the total error (yellow).}
\label{fig:error}
\end{center} 
\end{figure}

The excess definitions used in this work, in \citet{cruz2014} and in \citet{morales2012} provide the lower number 
of detections, which do not include all objects whose excesses are supported by {\em Spitzer} observations. These
are the only approaches that include the {\it WISE} flux calibration error in the computation of the uncertainty. 
As an example, in Fig.~\ref{fig:error} we illustrate the impact that this source has on the total uncertainty 
on the flux ratio  $f_{22}$/$f_{12}$  for the 1,222 stars with and without planets: calibration error is dominant for objects with $f_{22}\gtrsim$20~mJy, which is the case 
for all stars with {\em Spitzer} data. This fact seems to point out that the {\it WISE} calibration error may be 
overestimated.

In general, the stars which present IR excess in this work share similarities. In the case of HD218396 (HR8799), 
a well known planetary system with 5 planets, a debris disc with a cold and warm component is 
found \citep{ballering2013}. The same happens to the stars without planets HD85301 and recently to HD107146, 
where \citet{macgregor2016} add a cold component (30K) to its disc, in agreement to the {\em Spitzer} photometry 
that also shows the excesses for these stars \citep{chen2005}. Some other debris disc were already found by 
\citet{patel2014} (HD136544, HD60491, HD11938) where discs within 95K and 190 K are modeled. HD106906 is a 
spectroscopic binary, in the Lower Centaurus Crux stellar association, that hosts a well known debris disc, 
 recently observed with adaptive optics systems \citep{wu2016,lagrange2016}. Its strong 
excess in the W4 band is in perfect agreement with the {\it Spitzer} photometry of \citet{chen2005}, whose 
blackbody fit of the excess provides a temperature of 90~K \citep{chen2005}. Only BD-10\,3166 is not catalogued 
previously as a debris disc host. Furthermore, \citet{lodieu2014} report this star with a M5 star as a large 
proper motion companion.

We performed a Welch's t-test \citep{welch1947}, suitable for data populations with different variance, to 
quantify the statistical significance of the difference in the number of IR excess detections between the 
two sample of stars, with or without planets. We consider the results obtained by the five excess definitions 
analyzed in this work, using the ``corrected" {\it WISE} photometry. We also take into account two detection 
thresholds of 3$\sigma$ and 4$\sigma$. In no case the test indicates that the difference in the percentage of 
IR excess detections is statistically significant, assuming a confidence level of 95\%. 
Our analysis therefore indicates that the presence of warm debris desks (assuming that all detected excesses 
are caused by such an object) are not correlated to the presence of planets. 

Finally, it is important to note that the detection rate of warm debris desks of $\lesssim$1\% is much lower 
than that of cold debris discs, making them a rare phenomena. As \citet{wyatt2008} mentioned, in fact, the 
rare finding of warm debris discs in FGK MS stars is due to a combination of their long lifetimes and the 
short periods of time in which dust is produced by a transient event.

\section*{Acknowledgements}

The authors thank the referee Antonio Hales for comments that
improved the presentation of the paper. We would like to thank also Grant Kennedy for useful comments. 
RFM, MC, EB and FC want to thank CONACyT for financial support through grants CB-2011-169554 and CB-2015-256961. 
This research has made use of the SIMBAD database, operated at CDS, Strasbourg, France.

%%%%%%%%%%%%%%%%%%%%%%%%%%%%%%%%%%%%%%%%%%%%%%%%%%

%%%%%%%%%%%%%%%%%%%% REFERENCES %%%%%%%%%%%%%%%%%%

% The best way to enter references is to use BibTeX:

\bibliographystyle{mnras}
\bibliography{biblio2} % if your bibtex file is called example.bib

\begin{thebibliography}{}
\makeatletter
\relax
\def\mn@urlcharsother{\let\do\@makeother \do\$\do\&\do\#\do\^\do\_\do\%\do\~}
\def\mn@doi{\begingroup\mn@urlcharsother \@ifnextchar [ {\mn@doi@}
  {\mn@doi@[]}}
\def\mn@doi@[#1]#2{\def\@tempa{#1}\ifx\@tempa\@empty \href
  {http://dx.doi.org/#2} {doi:#2}\else \href {http://dx.doi.org/#2} {#1}\fi
  \endgroup}
\def\mn@eprint#1#2{\mn@eprint@#1:#2::\@nil}
\def\mn@eprint@arXiv#1{\href {http://arxiv.org/abs/#1} {{\tt arXiv:#1}}}
\def\mn@eprint@dblp#1{\href {http://dblp.uni-trier.de/rec/bibtex/#1.xml}
  {dblp:#1}}
\def\mn@eprint@#1:#2:#3:#4\@nil{\def\@tempa {#1}\def\@tempb {#2}\def\@tempc
  {#3}\ifx \@tempc \@empty \let \@tempc \@tempb \let \@tempb \@tempa \fi \ifx
  \@tempb \@empty \def\@tempb {arXiv}\fi \@ifundefined
  {mn@eprint@\@tempb}{\@tempb:\@tempc}{\expandafter \expandafter \csname
  mn@eprint@\@tempb\endcsname \expandafter{\@tempc}}}

\bibitem[\protect\citeauthoryear{{Adibekyan}, {Sousa}, {Santos}, {Delgado
  Mena}, {Gonz{\'a}lez Hern{\'a}ndez}, {Israelian}, {Mayor}  \&
  {Khachatryan}}{{Adibekyan} et~al.}{2012}]{adibekyan2012}
{Adibekyan} V.~Z.,  {Sousa} S.~G.,  {Santos} N.~C.,  {Delgado Mena} E.,
  {Gonz{\'a}lez Hern{\'a}ndez} J.~I.,  {Israelian} G.,  {Mayor} M.,
  {Khachatryan} G.,  2012, \mn@doi [\aap] {10.1051/0004-6361/201219401}, \href
  {http://adsabs.harvard.edu/abs/2012A%26A...545A..32A} {545, A32}

\bibitem[\protect\citeauthoryear{{Ballering}, {Rieke}, {Su}  \&
  {Montiel}}{{Ballering} et~al.}{2013}]{ballering2013}
{Ballering} N.~P.,  {Rieke} G.~H.,  {Su} K.~Y.~L.,   {Montiel} E.,  2013,
  \mn@doi [\apj] {10.1088/0004-637X/775/1/55}, \href
  {http://adsabs.harvard.edu/abs/2013ApJ...775...55B} {775, 55}

\bibitem[\protect\citeauthoryear{{Bertone}, {Buzzoni}, {Ch{\'a}vez}  \&
  {Rodr{\'{\i}}guez-Merino}}{{Bertone} et~al.}{2004}]{bertone2004}
{Bertone} E.,  {Buzzoni} A.,  {Ch{\'a}vez} M.,   {Rodr{\'{\i}}guez-Merino}
  L.~H.,  2004, \mn@doi [\aj] {10.1086/422486}, \href
  {http://adsabs.harvard.edu/abs/2004AJ....128..829B} {128, 829}

\bibitem[\protect\citeauthoryear{{Bertran de Lis}, {Delgado Mena}, {Adibekyan},
  {Santos}  \& {Sousa}}{{Bertran de Lis} et~al.}{2015}]{bertran2015}
{Bertran de Lis} S.,  {Delgado Mena} E.,  {Adibekyan} V.~Z.,  {Santos} N.~C.,
  {Sousa} S.~G.,  2015, \mn@doi [\aap] {10.1051/0004-6361/201424633}, \href
  {http://adsabs.harvard.edu/abs/2015A%26A...576A..89B} {576, A89}

\bibitem[\protect\citeauthoryear{{Bessell}}{{Bessell}}{1979}]{bessel1979}
{Bessell} M.~S.,  1979, \mn@doi [\pasp] {10.1086/130542}, \href
  {http://adsabs.harvard.edu/abs/1979PASP...91..589B} {91, 589}

\bibitem[\protect\citeauthoryear{{Casagrande}, {Sch{\"o}nrich}, {Asplund},
  {Cassisi}, {Ram{\'{\i}}rez}, {Mel{\'e}ndez}, {Bensby}  \&
  {Feltzing}}{{Casagrande} et~al.}{2011}]{casagrande2011}
{Casagrande} L.,  {Sch{\"o}nrich} R.,  {Asplund} M.,  {Cassisi} S.,
  {Ram{\'{\i}}rez} I.,  {Mel{\'e}ndez} J.,  {Bensby} T.,   {Feltzing} S.,
  2011, \mn@doi [\aap] {10.1051/0004-6361/201016276}, \href
  {http://adsabs.harvard.edu/abs/2011A%26A...530A.138C} {530, A138}

\bibitem[\protect\citeauthoryear{{Castelli} \& {Kurucz}}{{Castelli} \&
  {Kurucz}}{2003}]{castelli2003}
{Castelli} F.,  {Kurucz} R.~L.,  2003, in {Piskunov} N.,  {Weiss} W.~W.,
  {Gray} D.~F.,  eds,  IAU Symposium Vol. 210, Modelling of Stellar
  Atmospheres. p.~A20

\bibitem[\protect\citeauthoryear{{Chen}, {Jura}, {Gordon}  \&
  {Blaylock}}{{Chen} et~al.}{2005}]{chen2005}
{Chen} C.~H.,  {Jura} M.,  {Gordon} K.~D.,   {Blaylock} M.,  2005, \mn@doi
  [\apj] {10.1086/428607}, \href
  {http://cdsads.u-strasbg.fr/abs/2005ApJ...623..493C} {623, 493}

\bibitem[\protect\citeauthoryear{{Chen}, {Mittal}, {Kuchner}, {Forrest},
  {Lisse}, {Manoj}, {Sargent}  \& {Watson}}{{Chen} et~al.}{2014}]{chen2014}
{Chen} C.~H.,  {Mittal} T.,  {Kuchner} M.,  {Forrest} W.~J.,  {Lisse} C.~M.,
  {Manoj} P.,  {Sargent} B.~A.,   {Watson} D.~M.,  2014, \mn@doi [\apjs]
  {10.1088/0067-0049/211/2/25}, \href
  {http://adsabs.harvard.edu/abs/2014ApJS..211...25C} {211, 25}

\bibitem[\protect\citeauthoryear{{Costa}, {Canto Martins}, {Le{\~a}o}, {Lima},
  {Freire da Silva}, {de Freitas}  \& {De Medeiros}}{{Costa}
  et~al.}{2016}]{costa2016}
{Costa} A.~D.,  {Canto Martins} B.~L.,  {Le{\~a}o} I.~C.,  {Lima} Jr J.~E.,
  {Freire da Silva} D.,  {de Freitas} D.~B.,   {De Medeiros} J.~R.,  2016,
  preprint, \href {http://adsabs.harvard.edu/abs/2016arXiv161102250C} {}
  (\mn@eprint {arXiv} {1611.02250})

\bibitem[\protect\citeauthoryear{Cruz-Saenz~de Miera}{Cruz-Saenz~de
  Miera}{2012}]{cruz2012}
Cruz-Saenz~de Miera F.,  2012, Master's thesis, Instituto Nacional de
  Astrof{\'\i}sica, {\'O}ptica y Electr{\'o}nica

\bibitem[\protect\citeauthoryear{{Cruz-Saenz de Miera}, {Chavez}, {Bertone}  \&
  {Vega}}{{Cruz-Saenz de Miera} et~al.}{2014}]{cruz2014}
{Cruz-Saenz de Miera} F.,  {Chavez} M.,  {Bertone} E.,   {Vega} O.,  2014,
  \mn@doi [\mnras] {10.1093/mnras/stt1888}, \href
  {http://adsabs.harvard.edu/abs/2014MNRAS.437..391C} {437, 391}

\bibitem[\protect\citeauthoryear{{Cutri} \& {et al.}}{{Cutri} \& {et
  al.}}{2012}]{cutri2012}
{Cutri} R.~M.,  {et al.} 2012, VizieR Online Data Catalog, \href
  {http://adsabs.harvard.edu/abs/2012yCat.2311....0C} {2311}

\bibitem[\protect\citeauthoryear{{Cutri} \& {et al.}}{{Cutri} \& {et
  al.}}{2013}]{cutri2013}
{Cutri} R.~M.,  {et al.} 2013, VizieR Online Data Catalog, \href
  {http://adsabs.harvard.edu/abs/2013yCat.2328....0C} {2328}

\bibitem[\protect\citeauthoryear{{Eiroa} et~al.,}{{Eiroa}
  et~al.}{2013}]{eiroa2013}
{Eiroa} C.,  et~al., 2013, \mn@doi [\aap] {10.1051/0004-6361/201321050}, \href
  {http://adsabs.harvard.edu/abs/2013A%26A...555A..11E} {555, A11}

\bibitem[\protect\citeauthoryear{{Fischer} \& {Valenti}}{{Fischer} \&
  {Valenti}}{2005}]{valefisch2005}
{Fischer} D.~A.,  {Valenti} J.,  2005, \mn@doi [\apj] {10.1086/428383}, \href
  {http://adsabs.harvard.edu/abs/2005ApJ...622.1102F} {622, 1102}

\bibitem[\protect\citeauthoryear{{Gonzalez}}{{Gonzalez}}{1997}]{gonzalez1997}
{Gonzalez} G.,  1997, \mn@doi [\mnras] {10.1093/mnras/285.2.403}, \href
  {http://adsabs.harvard.edu/abs/1997MNRAS.285..403G} {285, 403}

\bibitem[\protect\citeauthoryear{{Gustafsson}, {Edvardsson}, {Eriksson},
  {J{\o}rgensen}, {Nordlund}  \& {Plez}}{{Gustafsson}
  et~al.}{2008}]{gustafson2008}
{Gustafsson} B.,  {Edvardsson} B.,  {Eriksson} K.,  {J{\o}rgensen} U.~G.,
  {Nordlund} {\AA}.,   {Plez} B.,  2008, \mn@doi [\aap]
  {10.1051/0004-6361:200809724}, \href
  {http://adsabs.harvard.edu/abs/2008A%26A...486..951G} {486, 951}

\bibitem[\protect\citeauthoryear{{Hales}, {Barlow}, {Drew}, {Unruh}, {Greimel},
  {Irwin}  \& {Gonz{\'a}lez-Solares}}{{Hales} et~al.}{2009}]{hales2009}
{Hales} A.~S.,  {Barlow} M.~J.,  {Drew} J.~E.,  {Unruh} Y.~C.,  {Greimel} R.,
  {Irwin} M.~J.,   {Gonz{\'a}lez-Solares} E.,  2009, \mn@doi [\apj]
  {10.1088/0004-637X/695/1/75}, \href
  {http://adsabs.harvard.edu/abs/2009ApJ...695...75H} {695, 75}

\bibitem[\protect\citeauthoryear{{Haywood}}{{Haywood}}{2001}]{haywood2001}
{Haywood} M.,  2001, \mn@doi [\mnras] {10.1046/j.1365-8711.2001.04510.x}, \href
  {http://adsabs.harvard.edu/abs/2001MNRAS.325.1365H} {325, 1365}

\bibitem[\protect\citeauthoryear{{Jarrett} et~al.,}{{Jarrett}
  et~al.}{2011}]{jarret2011}
{Jarrett} T.~H.,  et~al., 2011, \mn@doi [\apj] {10.1088/0004-637X/735/2/112},
  \href {http://adsabs.harvard.edu/abs/2011ApJ...735..112J} {735, 112}

\bibitem[\protect\citeauthoryear{{Kennedy} \& {Wyatt}}{{Kennedy} \&
  {Wyatt}}{2012}]{kennedy2012}
{Kennedy} G.~M.,  {Wyatt} M.~C.,  2012, \mn@doi [\mnras]
  {10.1111/j.1365-2966.2012.21621.x}, \href
  {http://adsabs.harvard.edu/abs/2012MNRAS.426...91K} {426, 91}

\bibitem[\protect\citeauthoryear{{K{\'o}sp{\'a}l} et~al.,}{{K{\'o}sp{\'a}l}
  et~al.}{2013}]{kospal2013}
{K{\'o}sp{\'a}l} {\'A}.,  et~al., 2013, \mn@doi [\apj]
  {10.1088/0004-637X/776/2/77}, \href
  {http://adsabs.harvard.edu/abs/2013ApJ...776...77K} {776, 77}

\bibitem[\protect\citeauthoryear{{Krivov}, {Reidemeister}, {Fiedler},
  {L{\"o}hne}  \& {Neuh{\"a}user}}{{Krivov} et~al.}{2011}]{krivov2011}
{Krivov} A.~V.,  {Reidemeister} M.,  {Fiedler} S.,  {L{\"o}hne} T.,
  {Neuh{\"a}user} R.,  2011, \mn@doi [\mnras]
  {10.1111/j.1745-3933.2011.01133.x}, \href
  {http://adsabs.harvard.edu/abs/2011MNRAS.418L..15K} {418, L15}

\bibitem[\protect\citeauthoryear{{Kuchner} et~al.,}{{Kuchner}
  et~al.}{2016}]{kuchner2016}
{Kuchner} M.~J.,  et~al., 2016, \mn@doi [\apj] {10.3847/0004-637X/830/2/84},
  \href {http://adsabs.harvard.edu/abs/2016ApJ...830...84K} {830, 84}

\bibitem[\protect\citeauthoryear{{Lagrange} et~al.,}{{Lagrange}
  et~al.}{2016}]{lagrange2016}
{Lagrange} A.-M.,  et~al., 2016, \mn@doi [\aap] {10.1051/0004-6361/201527264},
  \href {http://cdsads.u-strasbg.fr/abs/2016A%26A...586L...8L} {586, L8}

\bibitem[\protect\citeauthoryear{{Lodieu}, {P{\'e}rez-Garrido}, {B{\'e}jar},
  {Gauza}, {Ruiz}, {Rebolo}, {Pinfield}  \& {Mart{\'{\i}}n}}{{Lodieu}
  et~al.}{2014}]{lodieu2014}
{Lodieu} N.,  {P{\'e}rez-Garrido} A.,  {B{\'e}jar} V.~J.~S.,  {Gauza} B.,
  {Ruiz} M.~T.,  {Rebolo} R.,  {Pinfield} D.~J.,   {Mart{\'{\i}}n} E.~L.,
  2014, \mn@doi [\aap] {10.1051/0004-6361/201424210}, \href
  {http://adsabs.harvard.edu/abs/2014A%26A...569A.120L} {569, A120}

\bibitem[\protect\citeauthoryear{{MacGregor} et~al.,}{{MacGregor}
  et~al.}{2016}]{macgregor2016}
{MacGregor} M.~A.,  et~al., 2016, \mn@doi [\apj] {10.3847/0004-637X/823/2/79},
  \href {http://adsabs.harvard.edu/abs/2016ApJ...823...79M} {823, 79}

\bibitem[\protect\citeauthoryear{{Markwardt}}{{Markwardt}}{2009}]{markwidth2009}
{Markwardt} C.~B.,  2009, in {Bohlender} D.~A.,  {Durand} D.,   {Dowler} P.,
  eds,  Astronomical Society of the Pacific Conference Series Vol. 411,
  Astronomical Data Analysis Software and Systems XVIII. p.~251 (\mn@eprint
  {arXiv} {0902.2850})

\bibitem[\protect\citeauthoryear{{Martins} \& {Coelho}}{{Martins} \&
  {Coelho}}{2007}]{martins2007}
{Martins} L.~P.,  {Coelho} P.,  2007, \mn@doi [\mnras]
  {10.1111/j.1365-2966.2007.11954.x}, \href
  {http://adsabs.harvard.edu/abs/2007MNRAS.381.1329M} {381, 1329}

\bibitem[\protect\citeauthoryear{Massey}{Massey}{1952}]{massey1952}
Massey F.~J.,  1952, Annals of Mathematical Statistics, 23, 435

\bibitem[\protect\citeauthoryear{{Matthews}, {Krivov}, {Wyatt}, {Bryden}  \&
  {Eiroa}}{{Matthews} et~al.}{2014}]{matthews2014}
{Matthews} B.~C.,  {Krivov} A.~V.,  {Wyatt} M.~C.,  {Bryden} G.,   {Eiroa} C.,
  2014, \mn@doi [Protostars and Planets VI]
  {10.2458/azu_uapress_9780816531240-ch023}, \href
  {http://adsabs.harvard.edu/abs/2014prpl.conf..521M} {pp 521--544}

\bibitem[\protect\citeauthoryear{{Montesinos} et~al.,}{{Montesinos}
  et~al.}{2016}]{montesinos2016}
{Montesinos} B.,  et~al., 2016, \mn@doi [\aap] {10.1051/0004-6361/201628329},
  \href {http://adsabs.harvard.edu/abs/2016A%26A...593A..51M} {593, A51}

\bibitem[\protect\citeauthoryear{{Morales}, {Padgett}, {Bryden}, {Werner}  \&
  {Furlan}}{{Morales} et~al.}{2012}]{morales2012}
{Morales} F.~Y.,  {Padgett} D.~L.,  {Bryden} G.,  {Werner} M.~W.,   {Furlan}
  E.,  2012, \mn@doi [\apj] {10.1088/0004-637X/757/1/7}, \href
  {http://adsabs.harvard.edu/abs/2012ApJ...757....7M} {757, 7}

\bibitem[\protect\citeauthoryear{{Moro-Mart{\'{\i}}n}
  et~al.,}{{Moro-Mart{\'{\i}}n} et~al.}{2015}]{moromartin2015}
{Moro-Mart{\'{\i}}n} A.,  et~al., 2015, \mn@doi [\apj]
  {10.1088/0004-637X/801/2/143}, \href
  {http://adsabs.harvard.edu/abs/2015ApJ...801..143M} {801, 143}

\bibitem[\protect\citeauthoryear{{Neugebauer} et~al.,}{{Neugebauer}
  et~al.}{1984}]{neugebauer1984}
{Neugebauer} G.,  et~al., 1984, \mn@doi [\apjl] {10.1086/184209}, \href
  {http://adsabs.harvard.edu/abs/1984ApJ...278L...1N} {278, L1}

\bibitem[\protect\citeauthoryear{{Patel}, {Metchev}  \& {Heinze}}{{Patel}
  et~al.}{2014}]{patel2014}
{Patel} R.~I.,  {Metchev} S.~A.,   {Heinze} A.,  2014, \mn@doi [\apjs]
  {10.1088/0067-0049/214/1/14}, \href
  {http://adsabs.harvard.edu/abs/2014ApJS..214...14P} {214, 14}

\bibitem[\protect\citeauthoryear{{Pecaut}, {Mamajek}  \& {Bubar}}{{Pecaut}
  et~al.}{2012}]{peacut2012}
{Pecaut} M.~J.,  {Mamajek} E.~E.,   {Bubar} E.~J.,  2012, \mn@doi [\apj]
  {10.1088/0004-637X/746/2/154}, \href
  {http://adsabs.harvard.edu/abs/2012ApJ...746..154P} {746, 154}

\bibitem[\protect\citeauthoryear{{Rieke} \& {Lebofsky}}{{Rieke} \&
  {Lebofsky}}{1985}]{rieke1985}
{Rieke} G.~H.,  {Lebofsky} M.~J.,  1985, \mn@doi [\apj] {10.1086/162827}, \href
  {http://adsabs.harvard.edu/abs/1985ApJ...288..618R} {288, 618}

\bibitem[\protect\citeauthoryear{{Rieke} et~al.,}{{Rieke}
  et~al.}{2005}]{rieke2005}
{Rieke} G.~H.,  et~al., 2005, \mn@doi [\apj] {10.1086/426937}, \href
  {http://adsabs.harvard.edu/abs/2005ApJ...620.1010R} {620, 1010}

\bibitem[\protect\citeauthoryear{{Santos} et~al.,}{{Santos}
  et~al.}{2011}]{santos2011}
{Santos} N.~C.,  et~al., 2011, \mn@doi [\aap] {10.1051/0004-6361/201015494},
  \href {http://adsabs.harvard.edu/abs/2011A%26A...526A.112S} {526, A112}

\bibitem[\protect\citeauthoryear{{Sibthorpe}, {Ivison}, {Massey}, {Roseboom},
  {van der Werf}, {Matthews}  \& {Greaves}}{{Sibthorpe}
  et~al.}{2013}]{sibthorpe2013}
{Sibthorpe} B.,  {Ivison} R.~J.,  {Massey} R.~J.,  {Roseboom} I.~G.,  {van der
  Werf} P.~P.,  {Matthews} B.~C.,   {Greaves} J.~S.,  2013, \mn@doi [\mnras]
  {10.1093/mnrasl/sls002}, \href
  {http://adsabs.harvard.edu/abs/2013MNRAS.428L...6S} {428, L6}

\bibitem[\protect\citeauthoryear{{Sinclair}, {Helling}  \&
  {Greaves}}{{Sinclair} et~al.}{2010}]{sinclair2010}
{Sinclair} J.~A.,  {Helling} C.,   {Greaves} J.~S.,  2010, \mn@doi [\mnras]
  {10.1111/j.1745-3933.2010.00945.x}, \href
  {http://adsabs.harvard.edu/abs/2010MNRAS.409L..49S} {409, L49}

\bibitem[\protect\citeauthoryear{{Sousa} et~al.,}{{Sousa}
  et~al.}{2008}]{sousa2008}
{Sousa} S.~G.,  et~al., 2008, \mn@doi [\aap] {10.1051/0004-6361:200809698},
  \href {http://adsabs.harvard.edu/abs/2008A%26A...487..373S} {487, 373}

\bibitem[\protect\citeauthoryear{{Sousa}, {Santos}, {Israelian}, {Mayor}  \&
  {Udry}}{{Sousa} et~al.}{2011}]{sousa2011}
{Sousa} S.~G.,  {Santos} N.~C.,  {Israelian} G.,  {Mayor} M.,   {Udry} S.,
  2011, \mn@doi [\aap] {10.1051/0004-6361/201117699}, \href
  {http://adsabs.harvard.edu/abs/2011A%26A...533A.141S} {533, A141}

\bibitem[\protect\citeauthoryear{{Trilling} et~al.,}{{Trilling}
  et~al.}{2008}]{trilling2008}
{Trilling} D.~E.,  et~al., 2008, \mn@doi [\apj] {10.1086/525514}, \href
  {http://adsabs.harvard.edu/abs/2008ApJ...674.1086T} {674, 1086}

\bibitem[\protect\citeauthoryear{{Uzpen} et~al.,}{{Uzpen}
  et~al.}{2007}]{uzpen2007}
{Uzpen} B.,  et~al., 2007, \mn@doi [\apj] {10.1086/511736}, \href
  {http://adsabs.harvard.edu/abs/2007ApJ...658.1264U} {658, 1264}

\bibitem[\protect\citeauthoryear{{Valenti} \& {Fischer}}{{Valenti} \&
  {Fischer}}{2005}]{valenti2005}
{Valenti} J.~A.,  {Fischer} D.~A.,  2005, \mn@doi [\apjs] {10.1086/430500},
  \href {http://adsabs.harvard.edu/abs/2005ApJS..159..141V} {159, 141}

\bibitem[\protect\citeauthoryear{{Welch}}{{Welch}}{1947}]{welch1947}
{Welch} B.~L.,  1947, Biometrika, 34, 28

\bibitem[\protect\citeauthoryear{{Wittenmyer}, {Endl}, {Wang}, {Johnson},
  {Tinney}  \& {O'Toole}}{{Wittenmyer} et~al.}{2011}]{wittenmyer2011}
{Wittenmyer} R.~A.,  {Endl} M.,  {Wang} L.,  {Johnson} J.~A.,  {Tinney} C.~G.,
   {O'Toole} S.~J.,  2011, \mn@doi [\apj] {10.1088/0004-637X/743/2/184}, \href
  {http://adsabs.harvard.edu/abs/2011ApJ...743..184W} {743, 184}

\bibitem[\protect\citeauthoryear{{Wu} et~al.,}{{Wu} et~al.}{2016}]{wu2016}
{Wu} Y.-L.,  et~al., 2016, \mn@doi [\apj] {10.3847/0004-637X/823/1/24}, \href
  {http://cdsads.u-strasbg.fr/abs/2016ApJ...823...24W} {823, 24}

\bibitem[\protect\citeauthoryear{{Wyatt}}{{Wyatt}}{2008}]{wyatt2008}
{Wyatt} M.~C.,  2008, \mn@doi [\araa] {10.1146/annurev.astro.45.051806.110525},
  \href {http://adsabs.harvard.edu/abs/2008ARA%26A..46..339W} {46, 339}

\bibitem[\protect\citeauthoryear{{Wyatt}, {Clarke}  \& {Booth}}{{Wyatt}
  et~al.}{2011}]{wyatt2011}
{Wyatt} M.~C.,  {Clarke} C.~J.,   {Booth} M.,  2011, \mn@doi [Celestial
  Mechanics and Dynamical Astronomy] {10.1007/s10569-011-9345-3}, \href
  {http://adsabs.harvard.edu/abs/2011CeMDA.111....1W} {111, 1}

\makeatother
\end{thebibliography}

% Alternatively you could enter them by hand, like this:
% This method is tedious and prone to error if you have lots of references
%%\begin{thebibliography}{99}
%%\bibitem[\protect\citeauthoryear{Author}{2012}]{Author2012}
%%Author A.~N., 2013, Journal of Improbable Astronomy, 1, 1
%%\bibitem[\protect\citeauthoryear{Others}{2013}]{Others2013}
%%Others S., 2012, Journal of Interesting Stuff, 17, 198
%%\end{thebibliography}

%%%%%%%%%%%%%%%%%%%%%%%%%%%%%%%%%%%%%%%%%%%%%%%%%%

%%%%%%%%%%%%%%%%% APPENDICES %%%%%%%%%%%%%%%%%%%%%

%\appendix

%\section{Some extra material}

%If you want to present additional material which would interrupt the flow of the main paper,
%it can be placed in an Appendix which appears after the list of references.

%%%%%%%%%%%%%%%%%%%%%%%%%%%%%%%%%%%%%%%%%%%%%%%%%%

% Don't change these lines
\bsp	% typesetting comment
\label{lastpage}
\end{document}